\documentclass{imsart}
\RequirePackage{natbib}

\usepackage{amsmath,amssymb,bm}
\usepackage{times,parskip,epsf,epsfig,latexsym,rotating,amsmath,amssymb,natbib,bm,epstopdf,float,epsf}
\arxiv{math.ST/0000000}

\begin{document}

\begin{frontmatter}

\title{Bayesian registration of functions and curves}
\runtitle{Bayesian registration}

\begin{aug}
  \author{\fnms{Wen}  \snm{Cheng}\corref{}\ead[label=e1]{chengwen1985@gmail.com}},
  \author{\fnms{Ian L.} \snm{Dryden}\thanksref{t1}\ead[label=e2]{ian.dryden@nottingham.ac.uk}}
  \and
  \author{\fnms{Xianzheng}  \snm{Huang}%
  \ead[label=e3]{huang@stat.sc.edu}} %

 \thankstext{t1}{To whom correspondence should be addressed. 
We thank David Hitchcock and Huiling Le for their comments, and acknowledge
the support of a Royal Society Wolfson Research Merit Award and EPSRC grant EP/K022547/1.}

  \runauthor{Cheng et al.}

  \affiliation{The University of South Carolina and The University of Nottingham}

  \address{Department of Statistics,\\Le Conte College,\\
The University of South Carolina,\\Columbia, SC 29208, USA.\\ 
          \printead{e1,e3}}

  \address{School of Mathematical Sciences,\\
The University of Nottingham,\\
University Park,\\Nottingham, NG7 2RD, UK.\\
          \printead{e2}}

\end{aug}

\begin{abstract}
Bayesian analysis of functions and curves is considered, where warping and other geometrical transformations are often required for meaningful comparisons.  We focus on two applications involving the classification of mouse vertebrae shape outlines and the alignment of mass spectrometry data in proteomics. The functions and curves of interest are represented using the recently introduced square root velocity function, which  enables a warping invariant elastic distance to be calculated in a straightforward manner. We distinguish between various spaces of interest: the original space, the ambient space after standardizing, and the quotient space after removing a group of transformations. Using Gaussian process models in the ambient space and Dirichlet priors for the warping functions, we explore Bayesian inference for curves and functions. Markov chain Monte Carlo algorithms are introduced for simulating from the posterior, including simulated tempering for multimodal posteriors. We also compare ambient and quotient space estimators for mean shape, and explain their frequent similarity in many practical problems using a Laplace approximation. A simulation study is carried out, as well as shape classification of the mouse vertebra outlines and practical alignment of the mass spectrometry functions. 
\end{abstract}



\end{frontmatter}

\section{Introduction}
We consider statistical analysis of functions and curves where some form of
registration or time warping is of interest. We focus on two applications involving
classification of mouse vertebrae shape outlines in evolutionary biology and the alignment of mass spectrometry 
data in proteomics. Both applications require methods which can take account of arbitrary 
reparameterizations of the functions or curves of interest. 
In order to help choose appropriate methods and models we  first describe 
three different spaces of interest: the original space, the ambient space and the quotient 
space. The choice of space in which to specify the statistical model is important, as 
it determines what type of mean estimation and subsequent statistical analyses are carried out.
Our main contribution is to introduce a Bayesian approach to the analysis of functions and 
curves, which is demonstrated to be effective in the two applications. Inference is carried out
using Markov chain Monte Carlo simulation, and prior beliefs about the amount of  
time warping or registration are included as part of the model.

We wish to consider applications where the functions or curves of interest may not be in alignment. For example, 
in the study of growth curves of children it makes sense to consider a time warping of the curves
so that the curves match up in a biologically meaningful way. Children reach various stages of development 
such as puberty at different times, and so when comparing growth curves it is sensible to first align 
the curves in time and then compare the different heights and growth rates of the children using the time-warped curves 
\citep{Ramsli98}. The function registration problem 
has been considered by a large number of authors, including \citet{Kneigass92,Silverman95,Ramsli98,Kneipetal00} and \citet{Srivastavaetal11a}, 
among many others.  
Quantities such as a population mean function 
and a population covariance function can then be estimated in the space of curves after alignment. In addition to the amplitude 
variability of the functions post registration, it is also of interest to analyze the variability in the
registration transformations themselves, which is also known as phase variability.  
When analyzing curves in two or three dimensions we have additional potential invariances, 
such as translation, rotation and possibly scale invariance. 

\begin{figure}[htbp]\centering 
\begin{tabular}{cc}
\includegraphics[width=6cm,height=4.125cm]{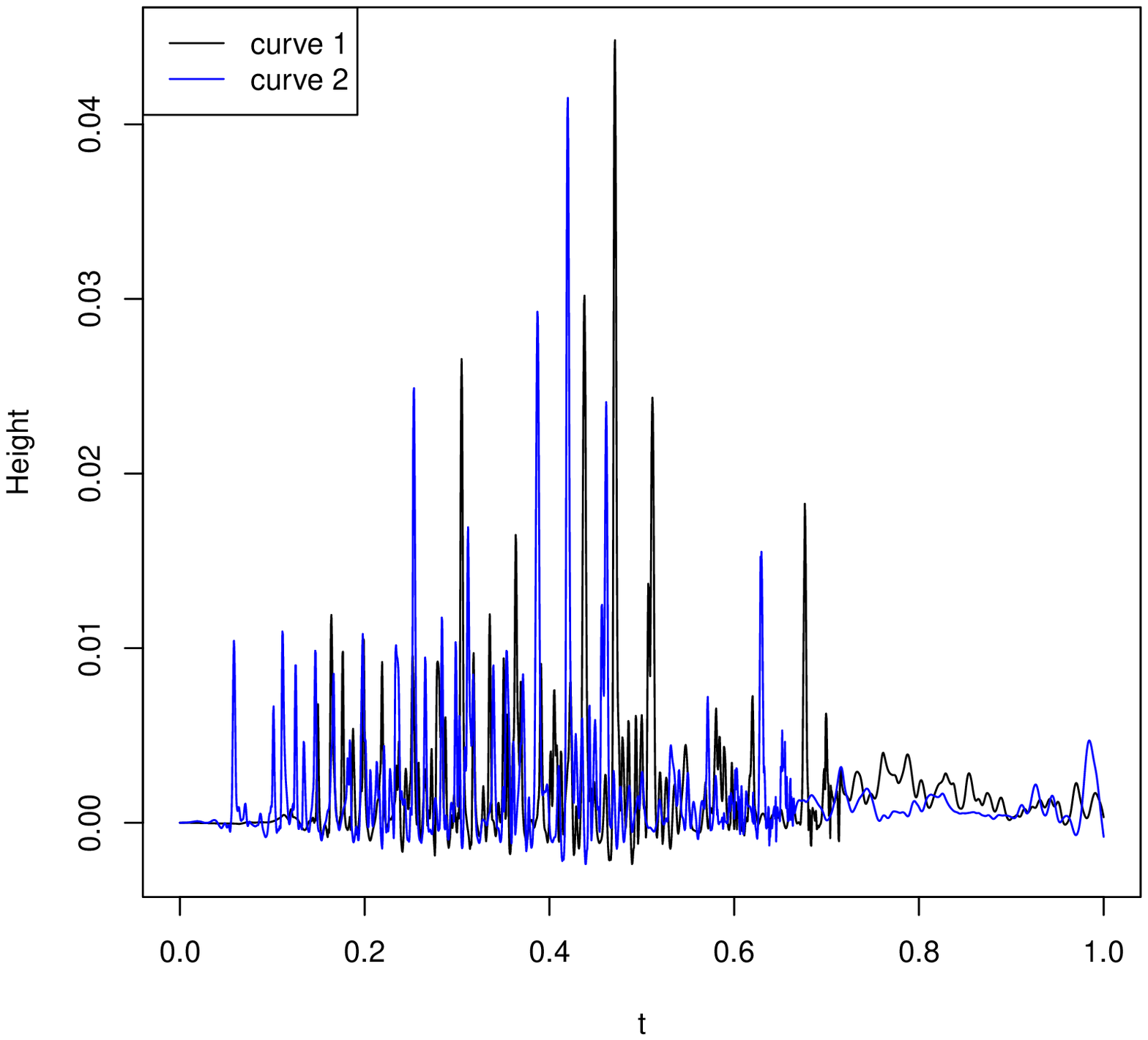} &
 \includegraphics[width=6cm,height=4.125cm]{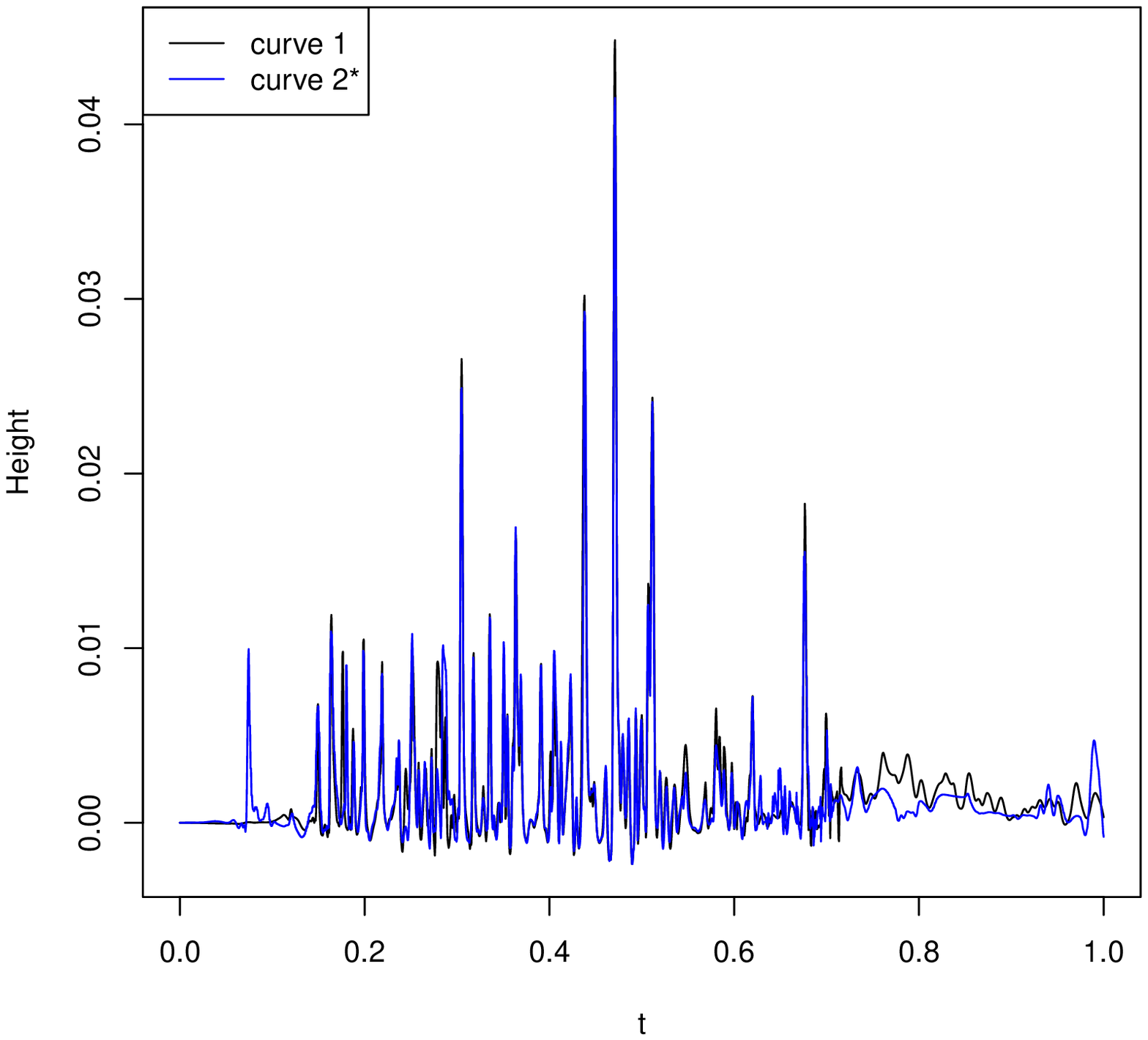}
\end{tabular}

\caption{Mass spectrometry scans, before registration (left) and after registration (right).} \label{fig3}
\end{figure}

As a motivating example consider the functions in Figure \ref{fig3}, which are two mass spectrometry scans from a larger dataset. 
In the left hand plot of Figure \ref{fig3} it can be seen that the scans are not well aligned, as 
the large peaks are not in the same positions in the x-axis. The goal of the alignment is to register the curves with a transformation 
of the x-axis so that peaks representing the same peptides can be compared between individuals.
After registration using the methodology of this paper it is clear in the right hand plot of Figure \ref{fig3} that all of the large peaks have been lined up. 
In this application it is suspected that much of the alignment can be accounted for by a translation of the x-axis, and so we develop a Bayesian method 
for alignment which can place strong prior information on the space of translations, if desired. The estimation of the alignments is 
obtained using the posterior mean of the warping functions, and inference is carried out using Markov chain Monte 
Carlo simulation. Further details are discussed in Section \ref{massspec5.2} after the methodology has been introduced, and we also consider 
a problem in shape analysis where it is of interest to classify vertebrae on the basis of the outline shape.  

\section{The spaces of interest} 
\subsection{Original, ambient and quotient spaces}
Consider data of interest in the form of functions or curves 
$$f_i(t) : [0,1] \rightarrow \mathbb{R}^m, i=1,\ldots,n.$$
In functional data 
analysis \citep{Ramssilv05} the function $f(t)$ is typically in $m=1$ 
dimension. In statistical shape analysis \citep{Klasetal03} the curve 
$f(t)$ is usually in $m=2$ or $m=3$ dimensions.  In practice we cannot 
observe a complete continuous function but rather a finite set of
discrete points $ \{ f(t_j) \in \mathbb{R}^m : j=1,\ldots,k \}$, where the function is observed 
at times $t_j, j=1,\ldots,k$.  

In a general form of the registration problem let us first consider the different spaces of interest. 
Each object $f$ is located in the original space 
(e.g. a space of functions, a space of curves in $\mathbb{R}^m$, or a space of landmark co-ordinates). 
The original space is where we represent the raw objects under study.

It is very common to standardize the objects with a preliminary transformation, such as 
centering or rescaling so that the objects have unit norm, or perhaps taking a derivative with respect to time to be translation invariant. 
These initial transformations 
are simple in nature and carried out individually on each object, very much in the spirit of standardizing
variables to have zero mean and unit variance in univariate statistics, or taking first differences in time series. 
The standardized object $f^*$  
is now represented in the ambient space $S$. Given that it is straightforward to transform to the ambient space, we will assume from
now on that this initial standardization has been carried out.

Finally we 
wish to investigate the equivalence class $[f] \in Q$ which is obtained by removing  transformations 
$\gamma \in G$ from the standardized $f^*$, where $G$ is a group of transformations and 
$Q = S / G$ is a quotient space. An important observation is that in order to compute distances in
the quotient space, {optimization} over the transformation group $G$ is required. 

This notion of equivalence class and quotient space is precisely that introduced by \citet{Kendall84} for 
the representation of the shapes of $k$ landmarks in $\mathbb{R}^m$, where $k > m$. The $k$ landmarks are points located in 
$m$ dimensions which represent the important features of the objects under study. 
In this situation the original space is 
the space of landmark co-ordinates $\mathbb R^{km} \setminus \{0\}$; the ambient space $S$ is the pre-shape sphere 
$S^{(k-1)m-1}$ of landmark coordinates which are Helmertized (or centered)  to remove location and  
scaled to have unit size;  and the quotient space is Kendall's shape space $\Sigma^k_m$ after quotienting out 
$G = SO(m) = \{ R : R^T R = RR^T = I_m, det(R)=1 \}$, where $SO(m)$ is the special orthogonal group of $m \times m$ rotation matrices.  
See \citet{Kendetal99} for a detailed description 
of the geometry of this space.

In functional data analysis the registration group is a transformation of the domain of the 
function, for example a translation $\gamma(t) = t+c$, affine transformation $\gamma(t) = at+c$, 
or the full group of diffeomorphic transformations $\gamma : [0,1] \rightarrow [0,1]$, 
such that $\gamma$ is 1-1 and onto. 
The functions themselves lie in the original space, are then standardized to the 
ambient space, and then finally are decomposed such that  
the amplitude variability is represented in the quotient space
and the phase variability is contained in the group of transformations $G$. 

In curve analysis the registration of interest is the transformation of the domain, and 
in addition we may wish to register using the translation, rotation and scaling of the curve. In this case the 
curves lie in the original space, standardized versions lie in the ambient space, then the shapes of the curves 
are represented in the quotient space. The main spaces used in this paper are given in Table \ref{examples}.

\begin{table}[htbp]
\hskip -1.5cm
\begin{center}
\begin{small}
\begin{tabular}{|cccc|}
\hline
Original object & Ambient space & Distance & Quotient space distance\\  
\hline
$X \in \mathbb{M}^{k \times m}$ & $Z = \frac{HX}{\| HX \|} \in S^{(k-1)m}$ &  
$\| Z_1 - Z_2 \|$ & ${\inf}_{\Gamma \in SO(m)} \| Z_1 - Z_2 \Gamma \|$ \\
 $\{ f(t) : t \in \mathbb{R} \}$ &   $q = \frac{\dot f}{| \dot f |^{1/2} } \in \mathbb{L}^2$ &  
$\| q_1 - q_2 \|_2$ &   ${\inf}_{\gamma \in G} \| q_1 - q_2 \circ \gamma \|_2$ \\
 $\{ f(t) : t \in \mathbb{R}^m \}$ &   $q = \frac{\dot f}{\| \dot f \|^{1/2} } \in \mathbb{L}^2$ &  
$\| q_1 - q_2 \|_2$ &   ${\inf}_{\gamma \in G,\Gamma \in SO(m)} \| q_1 - (q_2 \circ \gamma) \Gamma \|_2$ \\
\hline
\end{tabular}
\end{small}
\end{center}
\caption{Three examples of original objects, ambient spaces, ambient space distances and quotients spaces.
Row 1: $k$ landmarks in $m$ dimensions, where $H$ is a Helmert sub-matrix used for removing translation 
and $\Gamma$ is an $m \times m$ rotation matrix; row 2: 1-D functions, with warp $\gamma \in G$ a re-parameterization of time; 
row 3:   curves in $m$-D with warp $\gamma$ a re-parameterization of arc-length and $\Gamma$ is
a rotation matrix in $m$-dimensions.}\label{examples}
\end{table}

For our analysis of functions and curves, the original space and the 
ambient space $S$ are standard classical spaces, such as $\mathbb{L}^2$,  
$\mathbb{L}^2 \times \cdots \times \mathbb{L}^2$ or $S^{d-1}$, where statistical models 
can be relatively easily formulated, and inference carried out.  

In terms of statistical modelling and inference, working with objects in the 
group $G$ of transformations is more challenging, but can be undertaken. The 
geometry of the group is usually relatively simple and well understood.
However, the quotient space can be considerably more complicated in some situations. 
For example, the similarity shape space of a finite set of landmarks in 
three dimensions is very complicated, being a non-homogeneous space with 
singularities \citep{Lekend93}. 

So, an important question is: 
in which space shall we define our statistical model, the original, ambient or 
quotient space? Since the transformation from the original to the ambient space is 
quite straightforward, the main issue is whether we should consider models in the 
ambient space or the quotient space. Ultimately the choice of model will depend on the goals
of the study and what we are trying to make inference about. 

Let us first consider two data objects $X_1$ and $X_2$, 
which could both be standardized functions, curves, landmarks or any other type of object in 
an ambient space $S$. How close are $X_1$ and $X_2$, ignoring arbitrary registrations 
$\gamma_1, \gamma_2 \in G$? Let $[X_1]$ and $[X_2]$ denote the amplitudes (or shapes)
of $X_1, X_2$. 
A natural distance between the amplitude functions is in the quotient space: 
$$ d([X_1] , [X_2] ) = \inf_{\gamma \in G} d (X_1 , X_2 \circ \gamma ) , $$
where we must also have the isometric property 
$$ d([X_1 \circ \gamma^*] , [X_2 \circ \gamma^*] ) = d([X_1] , [X_2] ),$$
where an arbitrary common transformation $\gamma^*$ can be applied to both 
objects and the quotient distance remains unchanged. This property is 
also known as a parallel orbit property, in that the orbits (transformations
of an object by $\gamma^*$) are parallel, and it is also known as ``right-invariance''. 
This property is a necessity when thinking about practical statistical analyses which are
invariant to transformations.
If we apply an arbitrary transformation to our data then clearly all distances must remain invariant.

\subsection{Statistical models and inference}
Consider a distribution for a random object $X$, where it is the equivalence class
up to transformations in $\gamma \in G$ that is of interest. 
We have several choices for specifying a distribution. We could 
model $X$ in the ambient space with a population mean 
\begin{equation}
\mu_A = {\rm arg}\inf_{\nu \in S} \int_S d( x , \nu)^2 h(x) dx , \label{muA}
\end{equation}
where $h(x)$ is the probability density function (p.d.f.) of $X$. If $d(\cdot,\cdot )$ is the $\mathbb L^2$ or
Euclidean norm then $\mu_A = E[X] = \int x h(x) dx$. 
The key location parameter of 
interest is then the amplitude (shape) of $\mu_A$ written as $[\mu_A]$. 

Statistical models in the ambient space are quite straightforward to specify because the 
ambient space is usually not complicated.  
For example we specify a stochastic process/probability distribution for $X$, and then 
choose some coordinates in the quotient space, which we write as $U = [X]$ together with 
registration parameters $\gamma \in G$. We can specify a probability distribution for $X$ and 
transform from $X$ to $U$ (where $U = X \circ \gamma^{-1} \in Q$ and 
$\gamma \in G)$.  
Likelihood based inference about $\mu_A$ up to transformations $\gamma$ 
is then carried out after {marginalization}, i.e., after integrating out 
the transformations $\gamma$ from the distribution of $X$.  This approach was 
used by \citet{Marddryd89b,Drydmard91a,Drydmard92}
in landmark shape analysis for example.

Alternatively, we could model the equivalence class $U = [X]$ directly in the quotient space with 
population \cite{Frechet48} mean 
\begin{equation}
 \mu_Q = \mathop{{\rm arg}\inf_{\mu \in Q}}  \; \;  \int_Q d( u , \mu )^2 h(u) d u  , \label{muF}
\end{equation}
where $h(u)$ is the p.d.f. of $U$, and $d(\cdot,\cdot )$ is an intrinsic distance in the space. 
An intrinsic distance is the length of the shortest geodesic path 
between two points, where the path remains in the space at all times. 
The minimized value of the expected squared distance is known as the Fr\'echet variance, 
and we assume that a global minimum is obtained. If instead only a local minimum has been
found, we denote this as the Karcher mean \citep{Karcher77}.

Also, we could consider extrinsic distance between two points, where a space is embedded
in a higher dimensional Euclidean space. The extrinsic distance is taken as the Euclidean 
distance between the points in the embedding space. 
The population extrinsic mean 
\begin{equation}
 \mu_E = \mathop{{\rm arg}\inf_{\mu \in Q}}  \; \;   \int_Q d_E( u , \mu )^2 h(u) d u  ,  \label{muE}
\end{equation}
where $d_E(\cdot,\cdot )$ is an extrinsic distance. 
Models can be specified in the quotient space itself and we can
perform inference on $\mu_Q$ or $\mu_E$. 
The method requires {optimization} over the $\gamma$ parameters in order to compute
the intrinsic distances in the shape spaces. This is the approach used in 
Procrustes analysis \citep{Goodall91} in landmark shape analysis.

\begin{table}[htbp]
\begin{center}
\begin{tabular}{|ccc|}
\hline
Type of mean & Notation & Reference\\
\hline
Ambient space mean function& $\mu_A$ & Equation (\ref{muA})\\
Quotient space/Fr\'echet/Karcher mean function& $\mu_Q$ & Equation (\ref{muF})\\
Extrinsic mean function & $\mu_E$ & Equation (\ref{muE})\\
Ambient space mean vector& $\mu_A([t])$ & Section \ref{asymp}\\
Quotient space mean vector& $\mu_Q([t])$ & Section \ref{asymp}\\
\hline
\end{tabular}
\end{center}
\caption{Notation for the types of population means}\label{means}
\end{table}

A summary of the notation for the different types of population means considered in the paper is given in 
Table \ref{means}. 
In the next section we shall describe some  methods for computing distances and carrying out 
inference in quotient spaces 
for functions and curves. Then, in the following section we introduce our main approach 
to modelling using a Bayesian procedure in the ambient space.

\section{Quotient space}
\subsection{SRVF and quotient space}
Let $f$ be a real valued differentiable curve function in the original space, $f(t): [0,1]\rightarrow \mathbb R^m$. 
From \citet{Srivetal11} the Square Root Velocity Function (SRVF) of $f$  is defined as $q :[0,1]\rightarrow \mathbb R^m$, where 
$$q(t)=\frac{\dot{f}(t)}{\sqrt{\|\dot{f}(t)\|}} \; \; , $$ 
and $\|f\|$ denotes the standard Euclidean norm. 
After taking the derivative, the $q$ function is now invariant under translation of the original function, and is thus in the ambient space.
The main interests of this paper consider situations when $m=1$ for functions and $m=2$ for planar shapes. 
In the one dimensional functional case the domain $t \in [0,1]$ often represents `time' rescaled to unit length, whereas in two and higher dimensional cases 
 $t$ represents the proportion of arc-length along the curve.

Let $f$ be warped  by a re-parameterization $\gamma\in G$, i.e., $f\circ \gamma$, where $\gamma \in G$ : $[0,1]\rightarrow [0,1]$ is a strictly increasing 
differentiable warping function. The SRVF of $f\circ \gamma$ is then given as 
 $$q^*(t)=\sqrt{\dot{\gamma}(t)}q(\gamma(t)),$$
 using the chain rule. 
There are several reasons for using the $q$ representation instead of directly working with the original curve function $f$. One of the key reasons 
is that we would like to consider a metric that is invariant under re-parameterization transformation $G$. The 
elastic metric of \citet{Srivetal11} satisfies this desired property, 
$$d_{\rm Elastic}(f_1\circ\gamma,f_2\circ\gamma)=d_{\rm Elastic}(f_1,f_2),$$ 
although it is quite complicated to work with directly on the functions $f_1$ and $f_2$. 
However, the use of the SRVF representation 
simplifies the calculation of the elastic metric to an easy-to-use 
$\mathbb L^2$ metric between the SRVFs, which is attractive both theoretically and computationally.

If we define the group $G$ to be domain re-parameterization and we consider an equivalence class for $q$ functions under $G$, which is denoted as $[q]$, 
then we have the equivalence class $[q]\in Q$, where $Q$ is a quotient space after removing arbitrary domain warping. 
First consider the functional case in $m=1$ dimension. An elastic distance \citep{Srivetal11} defined in $Q$ is  given as the following   
$$d(q_1,q_2)=d([q_1],[q_2])=\inf_{\gamma\in G}\|q_1-\sqrt{\dot{\gamma}}q_2(\gamma)\|_2^2 = d_{\rm Elastic}(f_1,f_2),$$
where 
$\| q \|_2 = \{ \int_0^1 q(t)^2 dt \}^{1/2} $
denotes the $\mathbb L^2$ norm of $q$. For the $m=1$ dimensional case the elastic metric is equivalent to the Fisher-Rao 
metric for measuring distances between probability density functions. 
If $q_1$ can be expressed as some warped version of $q_2$, i.e., they are in the same equivalence class, then  $d([q_1],[q_2])=0$ in quotient space.
This elastic distance is a proper distance satisfying symmetry, non-negativity and the triangle inequality.
Note that we sometimes wish to remove scale from the function or curve, and hence we can standardize so that 
\begin{equation} 
\int_0^1 q(t)^2 dt = 1.  \label{unitsize}
\end{equation}  
In this case the ambient space would be the Hilbert sphere $S^{\infty}$.
 In the $m \ge 2$ dimensional case it is common to also require invariance under rotation of the original curve. Hence we may also wish to 
consider an elastic distance \citep{Joshietal07,Srivetal11} defined in $Q$ given as   
$$d([q_1],[q_2])=\inf_{\gamma\in G, \Gamma \in SO(m)}\|q_1-\sqrt{\dot{\gamma}}q_2(\gamma)\Gamma\|_2 .$$
The $m=2$ dimensional elastic metric for curves was first given by \citet{Younes98}.

\subsection{Quotient space inference}
Inference can be carried out directly in the quotient space $Q$, and in this case the population 
mean is most naturally the Fr\'echet/Karcher mean $\mu_Q$. Given a random sample $[q_1],\ldots,[q_n]$ we obtain the sample Fr\'echet mean by  
optimizing over the warps for the 1D function case \citep{Srivastavaetal11a}:
$${\hat\mu_Q} = {\rm arg}\inf_{\mu \in Q} \sum_{i=1}^{n} \inf_{\gamma_i \in G} \| \mu-\sqrt{\dot{\gamma_i}}(q_i\circ \gamma_i)\|_2^2 . $$
In addition for the $m \ge 2$ dimensional case \citep{Srivetal11} we also need to optimize over the rotation matrices $\Gamma_i$ where 
$${\hat\mu_Q} = {\rm arg}\inf_{\mu \in Q} \sum_{i=1}^{n} \inf_{\gamma_i \in G,\Gamma_i \in SO(m)} \| \mu-\sqrt{\dot{\gamma_i}}(q_i\circ \gamma_i)\Gamma_i\|_2^2.$$
This approach can be carried out using dynamic programming for pairwise matching, then 
ordinary Procrustes matching for the rotation, and the sample mean is given by 
$$ \hat\mu_Q = \frac{1}{n} \sum_{i=1}^n  \sqrt{  {\dot{\hat\gamma}}_i }  (q_i\circ \hat\gamma_i)\hat\Gamma_i . $$
Each of the parameters is then updated in an iterative algorithm 
until convergence.

\section{A Bayesian ambient space model}
\subsection{The likelihood for functions}
Our main approach is to consider a model in the ambient space, and then remove the unwanted transformations by marginalization. 
Since the $q$-function is a continuous function in the ambient space, naturally we consider a general 
stochastic process as the modelling framework for $q$, and we first consider the $m=1$ dimensional case. 
We assume a zero mean Gaussian process for the difference of two 1D $q$ functions, 
i.e., $\{ q_1-q^*_2 | \gamma \} \sim GP$, where $q_1$ is untransformed and $q_2^*$ is warped by a fixed reparameterization $\gamma$, i.e. 
$q^*_2(t)=\sqrt{\dot{\gamma}(t)}q_2(\gamma(t))$. The relative alignment function $\gamma$, contains the parameters of interest.

If we use $q_1([t])$ and $q^*_2([t])$ to denote $k+1$ finite points of $q_1(t)$ and $q^*_2(t)$ respectively, then the joint distribution of 
these $k$  finite differences is a multivariate normal distribution based on the Gaussian process assumption, i.e,  
$$\{ q_1([t])-q^*_2([t]) | \gamma \} \sim N_k(    0_k,    \Sigma_{k\times k}).$$ To simplify the problem, we 
assume $\Sigma_{k\times k}=\frac{1}{2\kappa}    I_{k\times k}$, where $\kappa$ is a concentration parameter,  
although more general covariance functions, such as the Gaussian or Mat\'ern functions, could be used.

\subsection{Prior distributions} 
If we treat the re-parameterization function $\gamma \in G$: $[0,1]\rightarrow [0,1]$ as a strictly increasing cumulative distribution function (c.d.f.), then this c.d.f. can be approximated by a set of equally spaced points along its domain $[0,1]$ and linear interpolation. Let $\gamma([t])$ denote $\{\gamma([t_i]),i=0, 1, 2, \dots, M\}$, the finite collection of $M+1$ discretized points and $[t_i]=\frac{i}{M}$, then we have $\gamma([t_0])=\gamma(0)=0$ and $\gamma([t_{M}])=\gamma(1)=1$. Further, if we let $p_{i}=\gamma([t_{i+1}])-\gamma([t_{ i}])$ for $i=1,2,\dots,M$, we have $0<p_i<1$ and $\sum_{i=1}^{M}p_i=1$. If we denote $\bm{p}_M =(p_1,p_2,\dots,p_M)$ and treat $\bm{p}_M$ as a random vector, we can assign a {\rm Dirichlet} prior to $\bm{p}_M |\gamma([t])$, i.e., $\pi(\bm{p}_M)\sim {\rm {
\rm Dirichlet}}(a_1,\ldots,a_M)$. We take equal $a_i = a$ here, writing ${\rm {
\rm Dirichlet}}(a)$. 
For $a=1$ the prior distribution is uniform and larger values of $a$ lead to transformations which are more concentrated on $\dot\gamma = 1$ (i.e. translations). In the limit as $M \to \infty$ the warping function is a {
\rm Dirichlet} 
process. The choice of $M$ is user specific, but it should be less than the number of discrete points in the $q$ functions, i.e. $M < k$.
The prior distribution for the concentration parameter $\kappa$ is taken as a Gamma($\alpha,\beta$) distribution, 
independent of $\gamma$. We use a fairly non-informative prior throughout with $\alpha = 1, \beta = 1,000$, and hence we have prior mean $E[\kappa ] = \alpha\beta = 1,000$
and prior variance $\alpha\beta^2 = 1,000,000$.

\subsection{Pairwise function comparison}
Combining the prior for $\gamma([t])$ and $\kappa$ with 
the likelihood model for finite differences of two $q$ functions, the posterior distribution for $\{\gamma([t]),\kappa\}$
given $(q_1([t]),q_2([t]))$ is 
$$\pi(\gamma,\kappa|q_1,q_2)\propto\kappa^{p/2}e^{-\kappa \| q_1([t])-\sqrt{\dot{\gamma}}(q_2([t])\circ \gamma)\|^2}\pi(\gamma)\pi(\kappa).$$ 

In the above model, $p$ represents the degrees of freedom in the model. 
If there is no unit scale length constraint (\ref{unitsize}) for $q$, then $p$ would be calculated as follows: $p=km$, where $k$ is the number of finite points taken from the $q$ function, and $m$ is the original function space dimension, i.e., $m=1$ for functions and $m \ge 2$ for curves in 
higher dimensions. One degree of freedom will be lost in the constrained case (\ref{unitsize}) and thus $p=km-1$.

In order to carry out inference on the warping function $\gamma$ and concentration parameter $\kappa$, we use
a Markov chain Monte Carlo (MCMC) algorithm to simulate from the joint posterior distribution. The concentration parameter 
$\kappa$  is updated using a Gibbs sampler as the conditional posterior for $\kappa$ given all other parameters is still Gamma distributed.  
For $\gamma([t])$ with $M+1$ points,  a shift in $\gamma([t_i])$ is
proposed at each discrete point $(i = 1, . . . ,M-1)$ and accepted/rejected according to a 
Metropolis-Hastings step. Note that $\gamma([t_0])=0$ and $\gamma([t_M])=1$ are both fixed and thus are not updated.
The resulting Markov chain is irreducible and aperiodic, and hence after a large number of iterations we have simulated dependent 
values from the posterior distribution.

\subsection{Multiple functions}
If we are interested in multiple functions or curves, we can specify a mean process 
for $q$ functions in the ambient space, i.e., $E(q^*_i)=\mu_A$, where $q^*_i=\sqrt{\dot{\gamma}_i}q_i(\gamma_i)$ is a warped version of $q_i$ through some underlying fixed $\gamma_i$. Based on the Gaussian process assumption again, we have
$$\{ q^*_i([t])-\mu_A([t])|\gamma_i([t]),\mu_A([t]) \} \sim N(    0_k,    \Sigma_{k\times k})$$
for $i=1,2,\dots,n$, where $n$ is the number of $q$ functions of interest.
We take the prior distribution of $\mu_A$ to be a zero mean Gaussian process with large variance, independent of all other parameters.
The joint posterior density for $(\mu_A,\gamma_1,\dots,\gamma_n)$ is then
$$\pi(\mu_A,\gamma_1,\dots,\gamma_n|q_1,\dots,q_n)\propto
\kappa^{np/2}e^{-\kappa\sum_{i=1}^{n} \| \mu_A([t])-q_i^*([t])\|^2}\pi(\mu_A)\pi(\gamma_1,\dots,\gamma_n)\pi(\kappa)$$ 
To simulate from the posterior distribution we again use an MCMC algorithm, consisting of pairwise MCMC updates from each curve to the 
current mean $\mu_A([t])$ and a Gibbs update for  $\mu_A([t])$ itself.

In order to compute the posterior mean estimate $\hat\mu_A([t])$ it is helpful to standardize in each MCMC iteration such that 
the Karcher mean of the warping functions from $\mu_A$ to each $q_i$ is the identity function, i.e. $\dot{\hat\gamma} = 1$.

\subsection{Curve Warping}
In the $m \ge 2$ dimensional case, we consider a Gaussian process for the difference of two vectorized $q$ functions in a relative orientation $\Gamma$, 
i.e., $\{ {\rm vec}(q_1-q^*_2)|\gamma,\Gamma\} \sim GP$, where $q^*_2=\sqrt{\dot{\gamma}}q_2(\gamma)\Gamma$. The matrix 
$\Gamma \in SO(m)$ is a rotation matrix with parameter vector $\theta$. If we assign a prior for rotation parameters (Eulerian angles) $\theta$ 
corresponding to rotation matrix $\Gamma$, then the joint posterior distribution of $(\gamma([t]),\theta)$, given $(q_1([t]),q_2([t]))$ is 
$$\pi(\gamma,\theta|q_1,q_2)\propto\kappa^{p/2}e^{-\kappa \| q_1([t])-q^*_2([t])\|^2}\pi(\gamma)\pi(\theta)\pi(\kappa),$$
where $\gamma, \theta, \kappa$ are independent {\it a priori}. 
Throughout the paper we take $\Gamma$ to have a Haar (uniform) prior on the space of rotation matrices. 
    
For the multiple curves case, define $q^*_i(t)=\sqrt{\dot{\gamma}_i(t)}q_i(\gamma_i(t))\Gamma_i$ and $\mu_A=E(q^*_i)$ for fixed $\gamma_i$ and $\Gamma_i$, and we assume $$\mu_A([t])-q^*_i([t])\sim N(0_{km},\Sigma_{km\times km})$$ for fixed $(\gamma_i,\Gamma_i)$, $i=1,\dots,n$. 
The joint posterior for  $(\mu_A,\gamma_1,\dots,\gamma_n,\Gamma_1,\dots,\Gamma_n)$ is  
$$\pi(\mu_A,\gamma_1,\dots,\gamma_n,\Gamma_1,\dots,\Gamma_n|q_1,\dots,q_n)\propto$$
$$\kappa^{np/2}e^{-\kappa\sum_{i=1}^{n} \| \mu_A-q_i^*\|^2} \pi(\mu_A)\pi(\gamma_1,\dots,\gamma_n)\pi(\Gamma_1,\dots,\Gamma_n)\pi(\kappa) ,$$
with warps, rotations and $\kappa$ independent {\it a priori}.  
Sampling from the posterior distribution is carried out through exactly the same procedure as when $m=1$ but 
with an extra Metropolis-Hastings update for  rotation angles.

\section{Properties}
\subsection{Asymptotic properties}\label{asymp}
Let us write $\mathbf{\phi}$ for the vector of all the
parameters in $\{ (\gamma_i, \Gamma_i), i=1,\ldots,n. \}$, and consider $\mu_A$ to be represented by a piecewise linear function connecting a finite 
number $k$ points given by $km$-vector $\mu_A([t])$.
The marginal posterior density for ambient space inference is given by  
\begin{equation}
\pi_A(\mu_A([t]),\kappa | X ) = \int_{\mathbf{\phi}} \pi(\mu_A([t]),\kappa,\mathbf{\phi} | X) d \mathbf{\phi}. \label{post0}
\end{equation}
The posterior mode estimator of $(\mu_A([t]),\kappa)$ is written as $(\hat\mu_A([t]),\hat\kappa)$ and is obtained by maximizing (\ref{post0}). If the
prior distribution of $(\mu_A([t]),\kappa)$ is uniform then $(\hat\mu_A([t]),\hat\kappa)$ is the maximum likelihood estimator. 
 If the prior is absolutely 
continuous in a neighbourhood of $\mu_A([t])$ with continuous positive density at $\mu_A([t])$ and the distribution satisfies certain regularity conditions 
(including differentiable in quadratic mean with non-singular Fisher information matrix $I_{\mu_A([t])}$), 
then consistency and asymptotic normality follow. 
Subject to the conditions of the Bernstein-von Mises theorem \citep[p141]{vanderVaart98}, we have  
$$ \sqrt{n}(\hat\mu_A([t]) - \mu_A([t])) \to N( \frac{1}{\sqrt{n}}\sum_{i=1}^n I_{\mu}^{-1} \dot \ell_{\mu_A([t])}(X_i) , I_{\mu_A([t])}^{-1})$$ 
in total variation norm as $n \to \infty$, 
where $\dot\ell_{\mu_A([t])}(X_i)$ is the derivative of the log-likelihood corresponding to observation $i$.
If $\hat\mu_A$ is a piecewise linear function obtained from the vector $\hat\mu_A([t])$, because $\hat\mu_A([t])$ is consistent for $\mu_A([t])$ 
we can equivalently state that
$ \hat\mu_A \to \mu_A$ 
in probability as $n \to \infty$, and hence the ambient space mean is consistent. \citet{Allaetal07} and \citet{Allaetal10} give detailed discussion of
consistency in ambient space models, in particular for deformable templates in image analysis.

The sample Fr\'echet mean vector $\hat\mu_Q([t])$ is consistent for the population Fr\'echet mean vector $\mu_Q([t])$ \citep{Kendall90,Le91} 
provided the distribution has support within a regular geodesic ball, and hence the corresponding piecewise linear function $\hat\mu_Q$ is consistent for $\mu_Q$.

\subsection{Comparison of the quotient and ambient space methods}
In general the population Fr\'echet  mean $\mu_Q$ in the quotient space and the ambient space mean $\mu_A$ do not have the same amplitude/shape, and hence the 
sample Fr\'echet mean can be inconsistent for the ambient space mean. Likewise the sample ambient space mean can be
inconsistent for the population Fr\'echet mean. 
It is most natural therefore to use the appropriate estimators given the choice of 
mean that is to be estimated. If we are interested in the amplitude/shape of the population ambient space mean $[\mu_A]$ then we use ambient space
inference, while if we are interested in the population Fr\'echet mean then we use the sample Fr\'echet mean. As we see below there are situations
where the sample ambient space and Fr\'echet estimators are very similar, and so our choice between them may be made on other grounds in this case, such as 
ease of computation.

When the prior distributions are uniform in the parameters an identical estimator to the sample Fr\'echet mean $\hat\mu_Q$
is obtained from the posterior mode in the Bayesian model of the previous section. If the priors are not uniform then
the posterior mode is in fact a penalized quotient estimator, with the objective function 
$$\hat\mu_{pen} = {\rm arg}\inf_{\mu \in Q} 
\sum_{i=1}^{n} \inf_{\gamma_i \in G,\Gamma_i \in SO(m)} \{ -\log\pi(\mu,\kappa, \gamma_i,\Gamma_i | q_1,\ldots,q_n ) \}  $$
for the curve case.

Note that in many practical situations the ambient space estimator and penalized quotient space estimators are quite similar. 
One reason for the similarity in practice is due to a Laplace approximation, and 
the marginal posterior density (for ambient space inference) is given by  
\begin{equation}
\pi_A(\mu,\kappa | X ) = \int_{\mathbf{\phi}} \pi(\mu,\kappa,\mathbf{\phi} | X) d \mathbf{\phi}. \label{post1}
\end{equation}
whereas the penalized quotient space estimator is obtained by maximization of  
\begin{equation}
\pi_Q( \mu,\kappa | X ) \propto \sup_{ \mathbf{\phi} }  \pi(\mu,\kappa, \mathbf{\phi} | X ) . \label{post2}
\end{equation}
where $X = \{ q_1,\ldots,q_n\}$.
Often we can consider $\pi_Q(\mu,\kappa | X )$ in (\ref{post2}) to be a good approximation to the marginal density (\ref{post1}) where the integral is
approximated using Laplace's method:
\begin{eqnarray}
\int_{\mathbf{\phi}} \pi(\mu,\kappa,\mathbf{\phi} | X) d \mathbf{\phi} & = & \int_{\mathbf{\phi}} 
b(\mathbf{\phi}) \exp\{ -A r(\mathbf{\phi}) \} d\mathbf{\phi} \nonumber\\
          & \approx & b(\hat{\mathbf{\phi}}) \left( \frac{2\pi}{A} \right)^{p/2} | \Sigma_{\hat{\mathbf{\phi}} } |^{1/2} 
\exp\{ -A r(\hat{\mathbf{\phi}} ) \} \nonumber \\
& \propto &  \sup_{ \mathbf{\phi} } \; \; \; b(\mathbf{\phi}) \exp\{ -A r( \mathbf{\phi} )\} \nonumber\\
& \propto & \pi_Q( \mu,\kappa | X ) . \nonumber
\end{eqnarray}
where the gradient of $r( \mathbf{\phi} )$ is zero at $\hat{ \mathbf{\phi} }$, $\Sigma_{\hat{ \mathbf{\phi} }}$ 
is the inverse of the positive definite Hessian matrix of  $r(\mathbf{\phi})$ at $\hat{\mathbf{\phi}}$ and $A$ is a constant.  
Laplace's approximation is exact when $(\mathbf{\phi}|\mu,\kappa)$ is multivariate Gaussian, i.e. 
$r(\mathbf{\phi})$ is a quadratic form in $\mathbf{\phi}$ and $b(\phi)$ is constant.

\subsection{Multimodality}\label{simtemp}
Multimodality of the posterior distribution can often be an issue with registration of functions and curves.
Simulated tempering \citep{Geyethom95} is a powerful simulation technique designed to overcome problems in moving between local modes 
of the posterior. The key idea is to first jump 
from the ``cold" temperature (target distribution), where it is difficult to move out of a local mode to a 
``hot" temperature where movement between modes is easier 
and then jump back to the ``cold" temperature. Using this procedure, the MCMC algorithm 
can explore the sample space in a more efficient manner. 

Let $\pi(\omega)\propto e^{-U(\omega)}$ be the unnormalized density which is the so called ``cold" distribution, where $\omega$ 
is the parameter vector. Often $\pi(\omega)$ 
has multiple local modes 
when the dimension of $\omega$ is high. In order to jump out of local modes in the updating algorithm, 
we need to make larger moves in the sampling space. Let $\pi_i(\omega)$ be a sequence of $T$ unnormalized 
densities where $\pi_i(\omega)\propto \pi(\omega)^{\beta_i}$ for $0\leq \beta_i<1$ and $i=1,\ldots,T$. Following \citet{Liu01}  
and \citet{Grametal10}   
$\beta_i$ is taken as:
$$\beta_i = (1+\delta_\beta)^{1-i}$$ which is geometric spacing with $\delta_\beta>0$.
The simulated tempering algorithm is then given as follows \citep{Geyethom95}:
\begin{itemize}
\item Given $\pi_i(\omega)$ update $\omega$ using a Metropolis-Hastings step or Gibbs step.
\item Generate $j=i\pm1 $ using probabilities $q_{i,j}$, where $q_{1,2}$=$q_{T,T-1}$=1 and $q_{i,i+1}=q_{i,i-1}=\frac{1}{2}$ if $1<i<T$.
\item Accept the proposal with probability $\min(r,1)$ where 
$$r = \frac{\pi_j(\omega)w_jq_{j,i}}{\pi_i(\omega)w_iq_{i,j}}.$$
\end{itemize}
Note that $w_j$ is the prior weight related to $\pi_j(\omega)$ such that each $\pi_j(\omega)$ is explored uniformly, i.e., the MCMC algorithm spends an 
equal amount of time in each of the $T$ densities.
In practice, the use of simulated tempering requires much tuning, and we use a simple strategy 
where the number of chains to run is $T=10$ and the spacing parameter $\delta_\beta=\frac{1}{\sqrt{N_T}}$ is used where $N_T$
is chosen such that the acceptance rate among jumping across chains is controlled to be roughly between 20\% to 40\%. 
Note that the $w_i$ need to be approximated from a preliminary run in which all $w_i$ are set equal. Based on the 
preliminary run, the $w_i$ is estimated to be $w_i \propto 1/n_i$ where $n_i$ is the number of samples that the MCMC algorithm takes from chain $i$. 
The number of iterations in the tuning pre-run is taken as $50,000$. In case any $n_i$ are equal to 0,  $\delta_\beta$ is decreased to 
$\delta_\beta=\frac{1}{K\sqrt{N_T}}$ with $K=2,3,\dots$ until all $n_i > 0$. 
The sampling of $\kappa$ is straightforward at each level, via a Gibbs sampler
$$\pi(\kappa|\gamma,q_1,q_2)\propto \Gamma(k_i(\frac{p}{2}+\alpha)+1-k_i,k_i(\beta+||q_1-\sqrt{\dot{\gamma}}(q_2\circ \gamma)||^2)),$$
and the sampling of $\gamma$ is via
$$\pi^{k_i}(\gamma|\kappa,q_1,q_2)\propto \{e^{-\kappa||q_1-\sqrt{\dot{\gamma}}(q_2\circ \gamma)||^2}\pi(\gamma) \}^{k_i} .$$

\section{Simulations and applications}
\subsection{Simulation Study}
 We consider now a simulation study to compare estimation properties of the quotient and ambient space estimators. The quotient space 
estimator $\hat{\mu}_Q$ is obtained by minimizing $\Sigma_{i=1}^{n} \| \mu-\sqrt{\dot{\gamma_i}}(q_i\circ \gamma_i)\|_2^2$ using 
dynamic programming while the ambient space estimator  $\hat\mu_A$ is obtained using the point-wise mean of posterior samples from MCMC iterations after convergence.

In a single Monte Carlo simulation repetition, a sample of $q$-functions in one dimension is generated through 
the model $q_i([t])=\sqrt{\dot{\gamma}_i}\mu_A(\gamma_i([t]))+e_i([t])$, where $e_i\sim N(0_k,\Sigma_{k\times k}), \Sigma = \sigma^2I_{k\times k}$ and 
$ \gamma_i\sim 
{\rm Dirichlet}(1)$ for $i=1,\ldots,n$.  

Both $\hat{\mu}_Q$ and $\hat\mu_A$ are computed and their Fisher-Rao distances to the 
underlying true $\mu_A$ are calculated. Note that since the goal is to estimate $\mu_A$ in the ambient space, it is expected that $\hat\mu_A$ 
will be more appropriate than $\hat\mu_Q$. The MCMC algorithm for $\hat\mu_A$ is run for $50,000$ iterations with a $25,000$ iteration burn-in period. 
The prior for $\gamma$ in the Bayesian model is taken as {\rm Dirichlet} with $a=1$, i.e. uniform. Given specific combinations of sample size $n \in \{5, 10, 20, 30, 
50, 100, 200 \}$ and error standard deviation $\sigma \in \{ 0.1, 0.3, 0.5, 1 \}$, 100 Monte Carlo 
repetitions are run and the arithmetic means of squared Fisher-Rao distances from 
both estimators to $\mu_A$ are recorded. 

\begin{figure}[htbp]\centering
\begin{tabular}{cccc} 
\includegraphics[width=4cm]{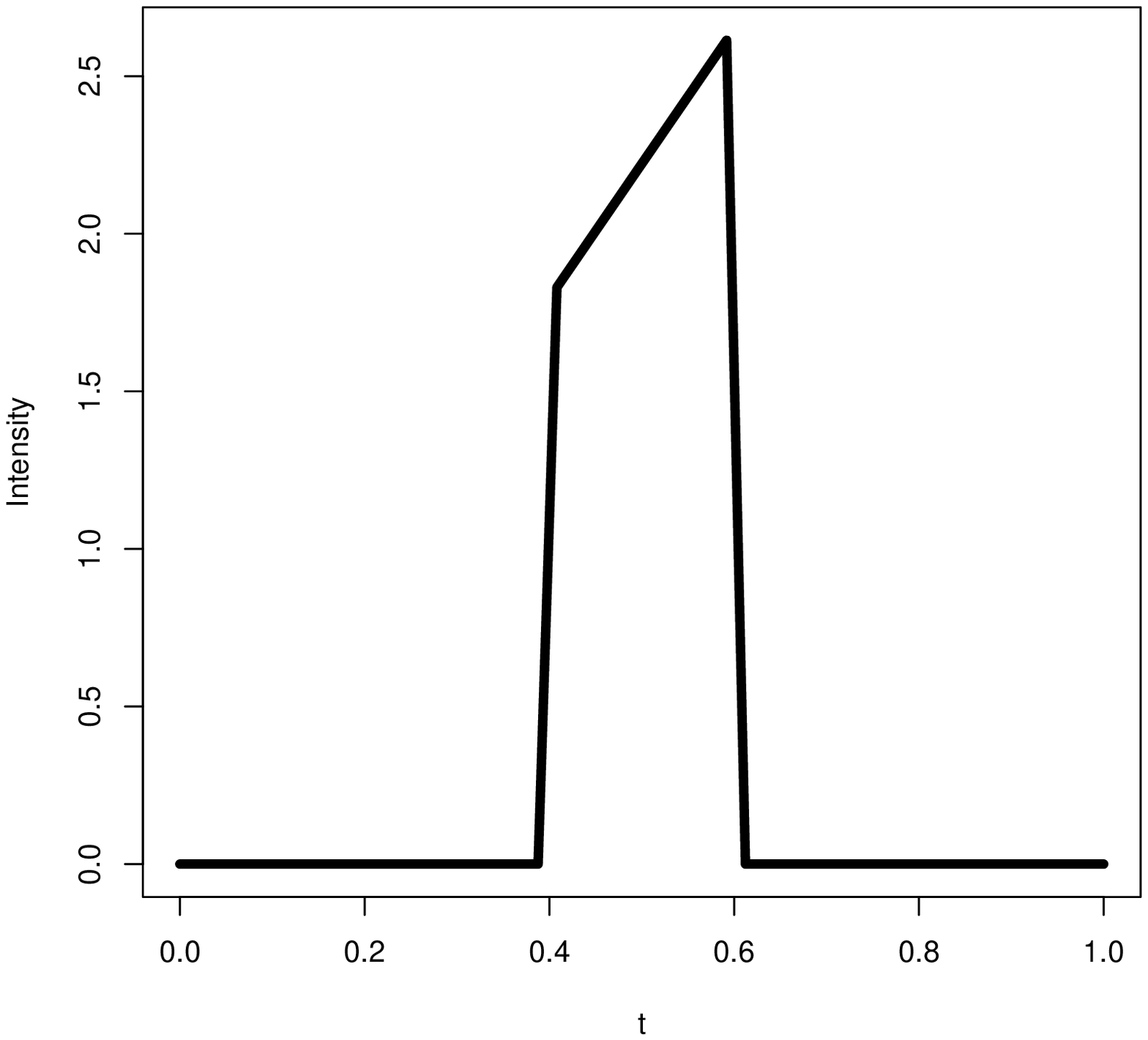}
\includegraphics[width=4cm]{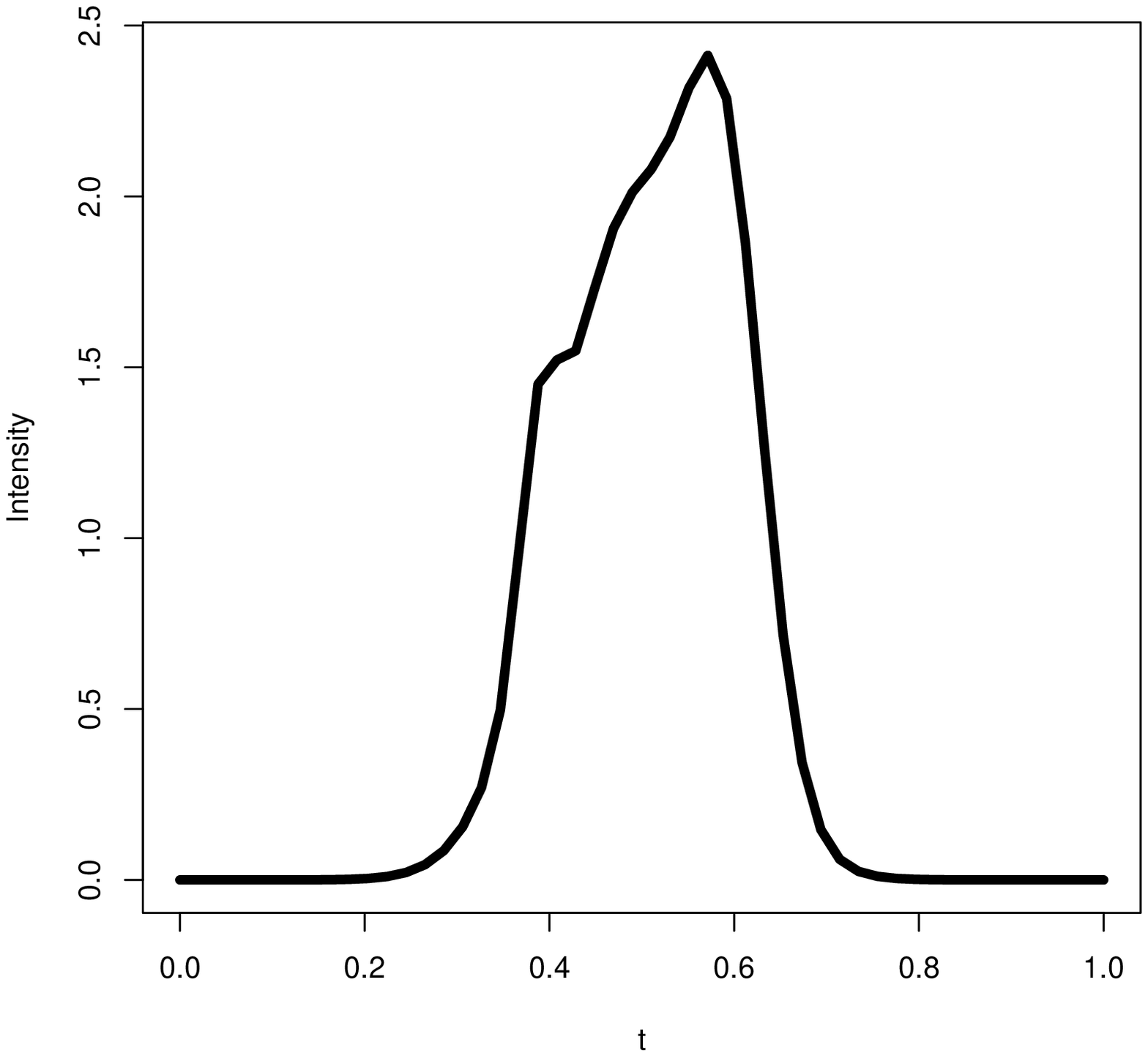}\\
\includegraphics[width=4cm]{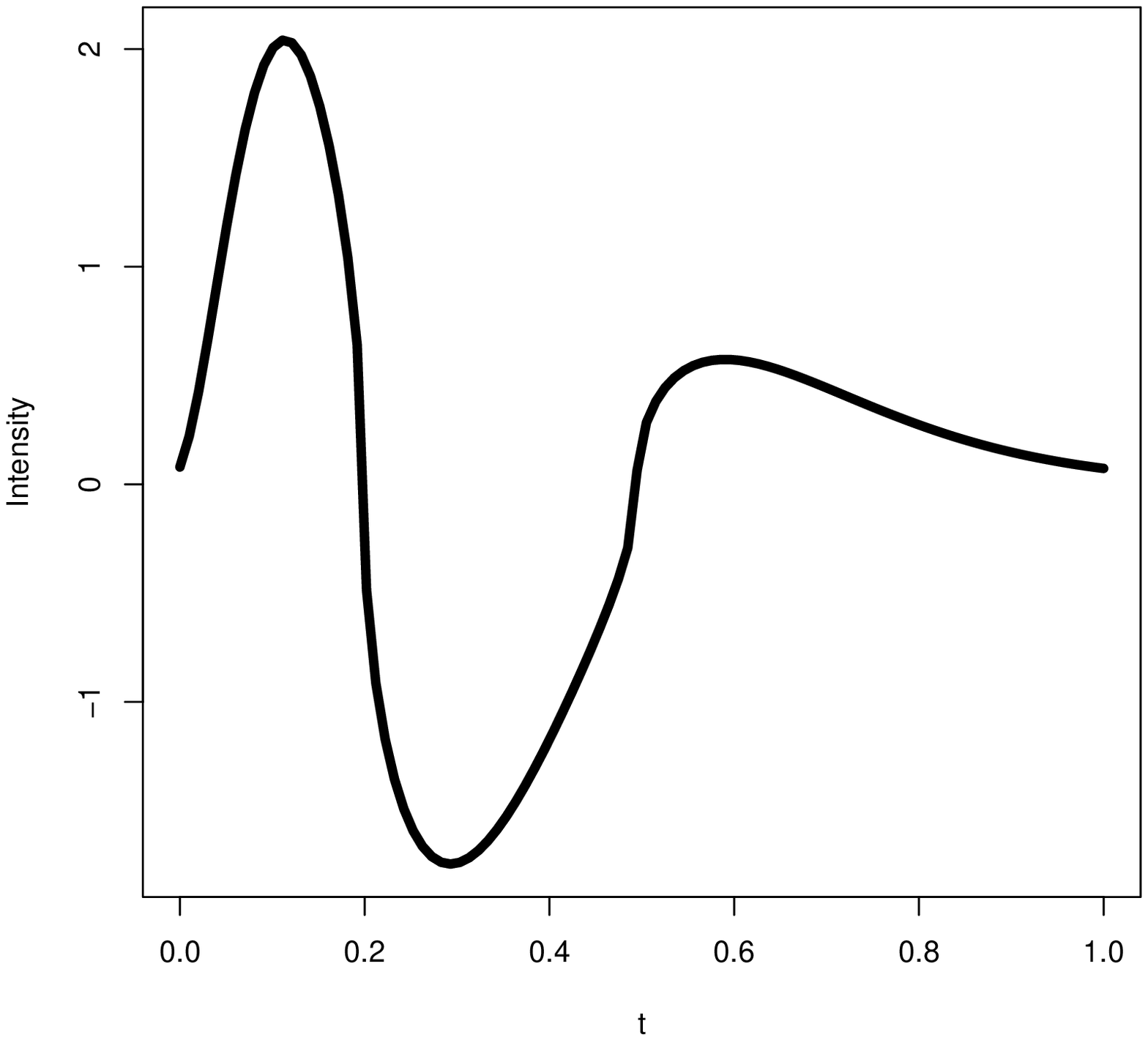}
\includegraphics[width=4cm]{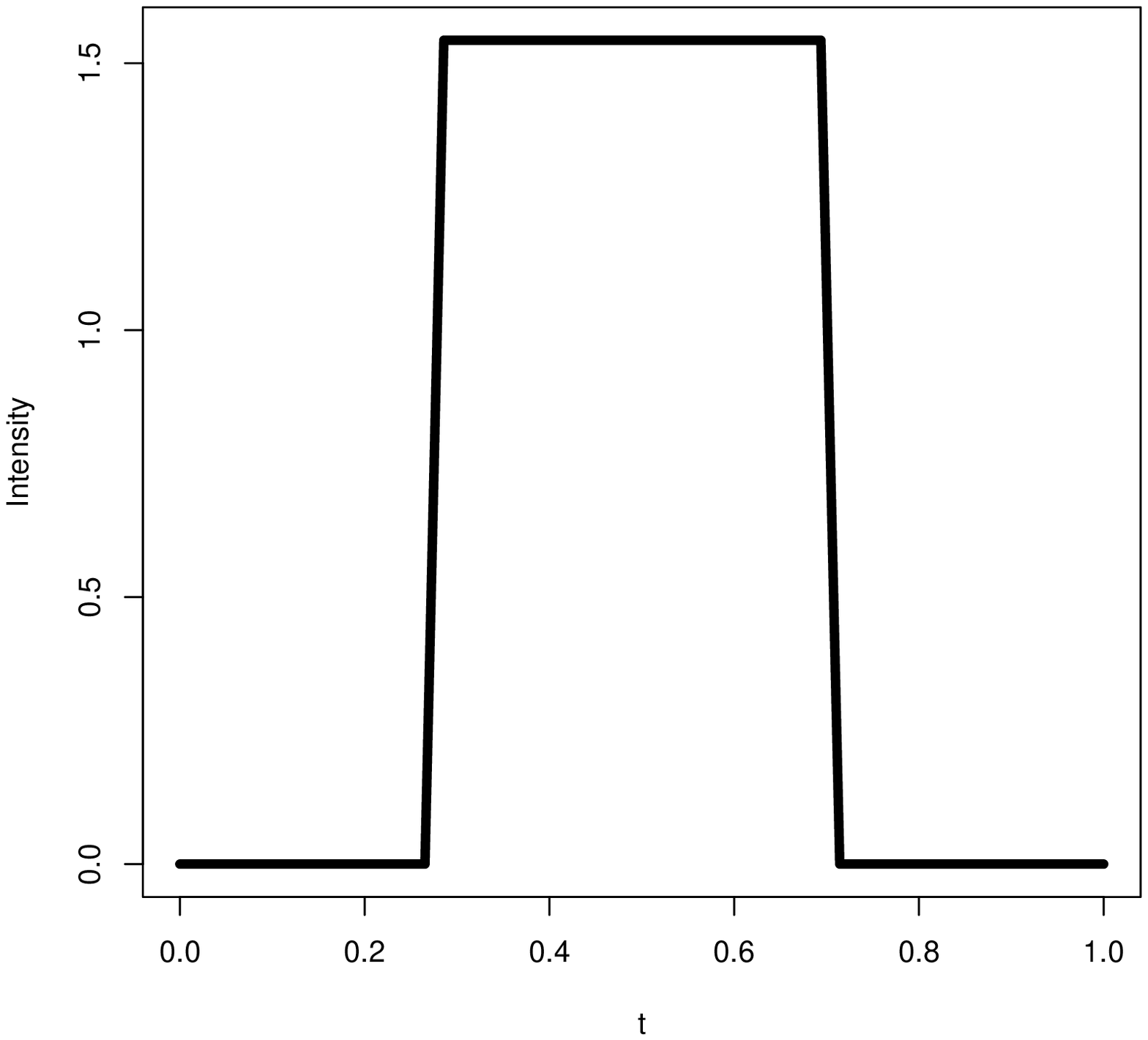}
\end{tabular}
\caption{The true $\mu_A(t)$ functions used for simulation study. From left to right we denote the 
functions as Example I,II,III and IV respectively. } \label{other}
\end{figure}

\begin{figure}[htbp]

\begin{center}
\includegraphics[width=12cm,angle=270]{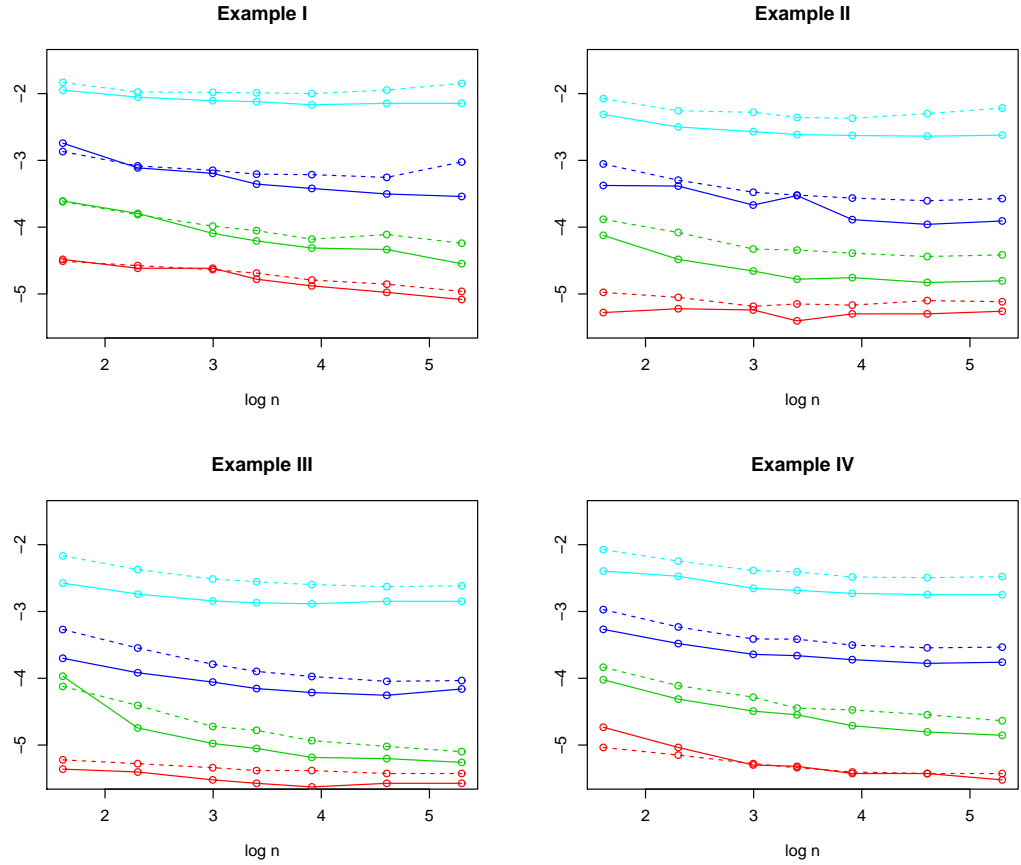}
\end{center}
\caption{The logarithm of the mean square Fisher Rao distance to the true mean $\mu_A$ versus logarithm of sample size $n$. 
The full line is the ambient space estimator and the dotted
line is the quotient space estimator. The colours are red ($\sigma=0.1$), green ($\sigma=0.3$),
blue ($\sigma=0.5$) and  cyan ($\sigma=1$).
}\label{FIGres}
\end{figure}

Four examples for $\mu_A$ are considered, which are all scaled to have unit length and unit time. The functions $\mu_A$ 
in examples I,II,III,IV given in Figure  \ref{other} are evaluated at $k$ equal to $51,51,101,51$ points respectively, 
and the warping functions are parameterized using $M+1$ points, where $M$ is equal to $10,10,20,10$ respectively. 
The underlying $\mu_A$ functions in example I and IV are piecewise linear, example II is a mixture of three normal
densities, and example III is the derivative of the difference of two Gamma functions (in fact it is 
the derivative of the canonical haemodynamic response function often used
to model the blood oxygen level dependent signals in fMRI \citep{Glover99}). 
The corresponding distances from both estimators are given in Figure \ref{FIGres}.

From Figure \ref{FIGres} we that see that when $\sigma$ is smaller, the average squared distance between the estimate and 
true value is smaller, and as $n$ increases in general the average squared distance becomes smaller. When $\sigma$ is 
small ($0.1$), the performance of both estimators is almost equivalent. 
However, for larger $\sigma$ in nearly all cases there is an advantage in using the ambient space estimator. One possible 
explanation could be over-warping of the quotient estimator to the noisy data due to the optimization over warpings, 
compared to the integration over warpings in the ambient space estimator. 
For large $\sigma \ge 0.5$ both procedures are clearly
biased for these values of $n$, but it must be borne in mind
that the signal to noise ratio is very low in these cases and so the estimation is very challenging and the discrete implementation 
will have an important effect.  
Overall, from these examples it does seem that there is an advantage in using the ambient space estimator as we expect, 
although this is at the expense of at least twice the computational time. 

\subsection{2D Mouse vertebrae}
A two-dimensional application is the study of the shape of the second thoracic (T2) vertebrae in mice 
\citep{Drydmard98}. Three groups of mouse vertebrae are available: 30 Control, 23 Large and 23 Small bones. The 
Large and Small group underwent genetic selection for large/small body weight, whereas the Control 
group consists of unselected mice. Each bone is represented by a curve consisting of 60 points which are 
determined through a semi-automatic procedure. Six landmarks are placed at points of 
high absolute curvature and then nine pseudo-landmarks are equally-spaced inbetween each pair of landmarks. 
The main interests here include carrying out pairwise registration, obtaining mean shapes and 
credibility intervals, and carrying out classification based on the registered shapes. It is very common in many application areas 
classify objects using shape information \citep{Drydmard98}, and for example in studying the fossil record there is a need to classify bones
from individuals into groups using size and/or shape as there is usually little or no other information available.  

\begin{figure}[htbp]\centering 
\begin{tabular}{c}
\includegraphics[height=5cm,width=5cm]{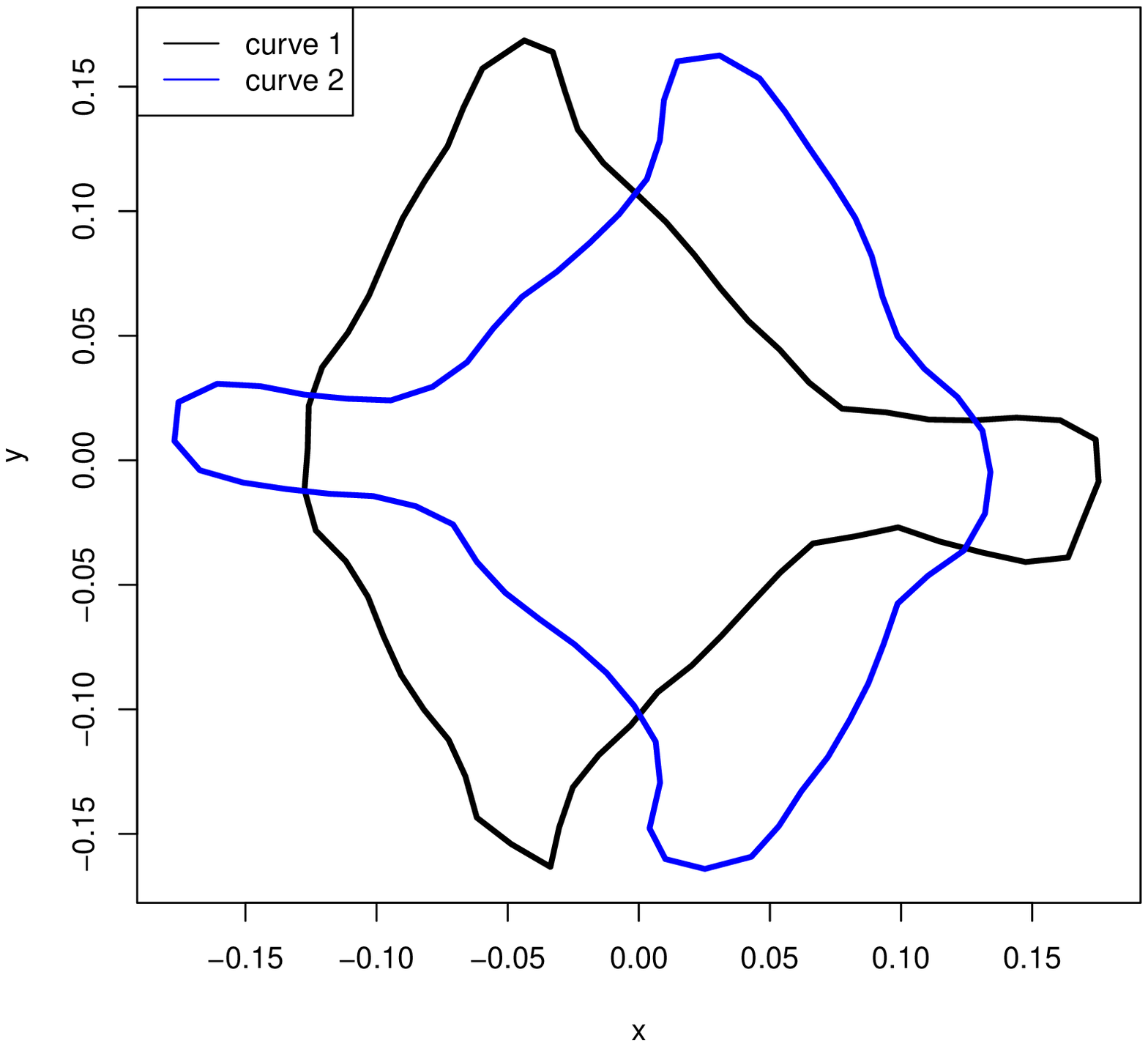} 
\includegraphics[height=5cm,width=5cm]{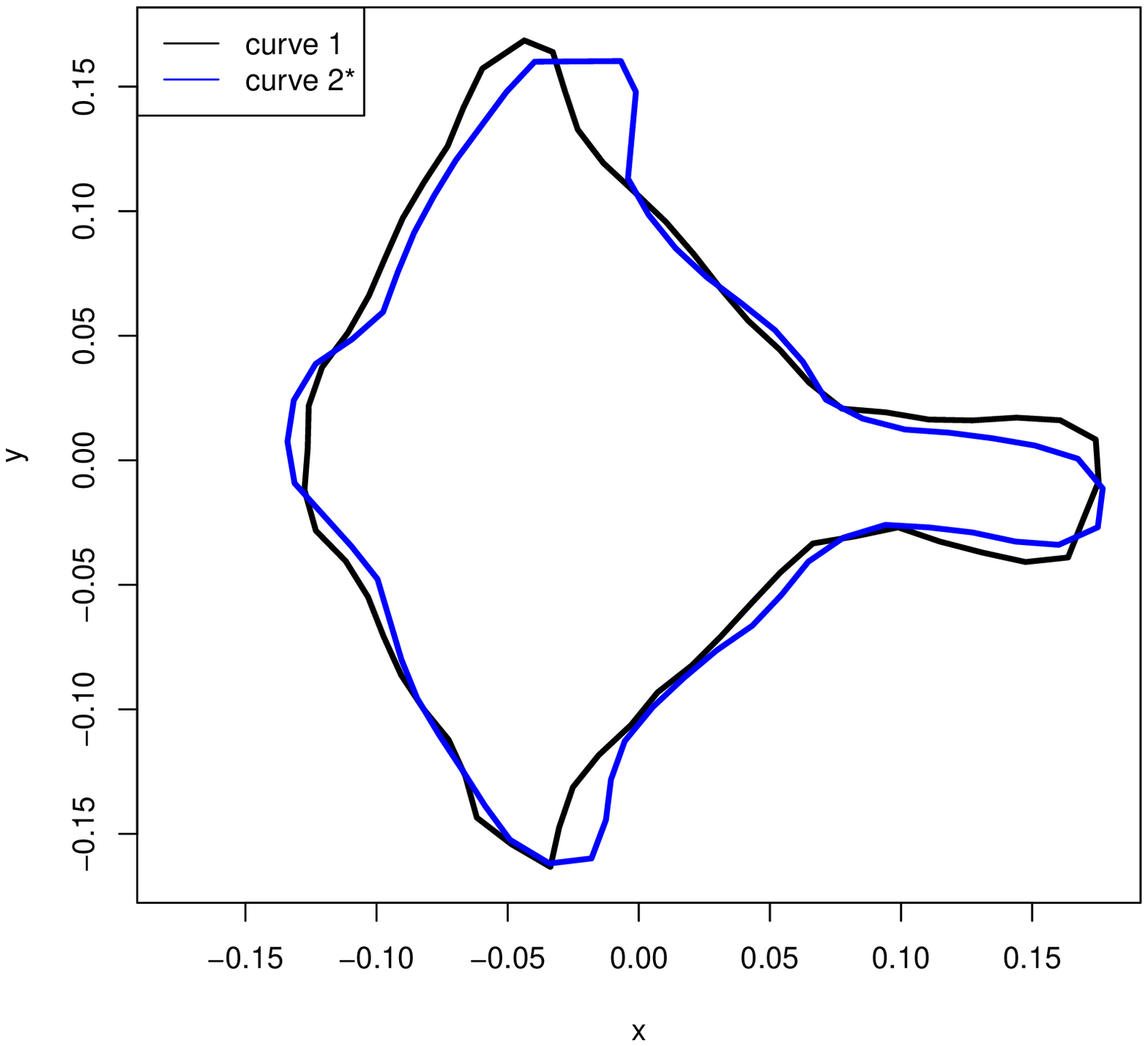} 
\end{tabular}
\caption{Unregistered curves on left and registration through $\hat\gamma(t)_A$ on right.}\label{pair}
\end{figure}

\begin{figure}[htbp]\centering 
\begin{tabular}{c} 
\includegraphics[height=5.5cm,width=5.5cm]{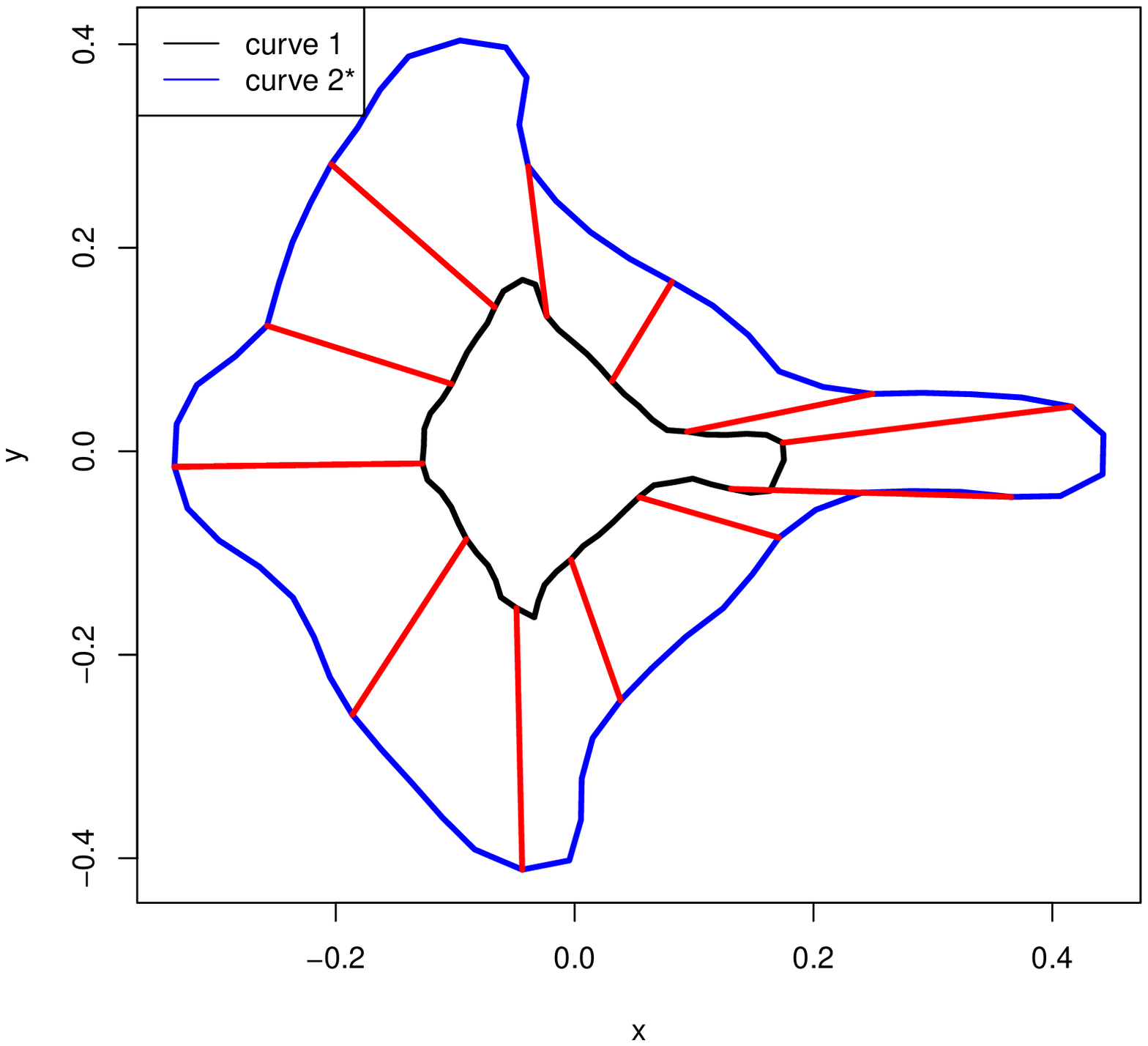}  
\includegraphics[height=5.5cm,width=5.5cm]{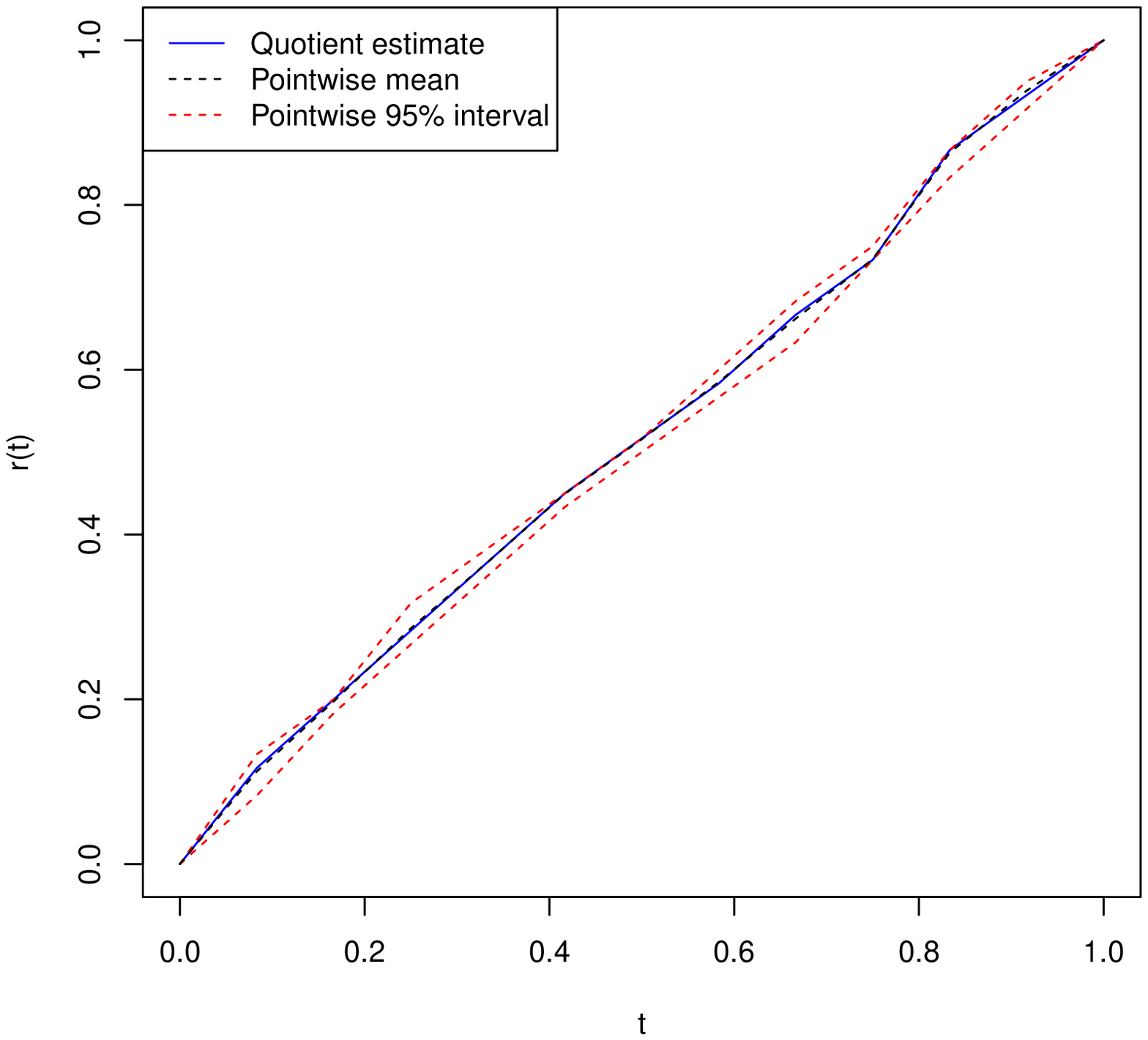} 
\end{tabular}
\caption{Correspondence based on $\hat\gamma(t)_A$ and 95\% credibility interval for $\gamma(t)$. One of the bones is drawn artificially smaller in order 
to better illustrate the correspondence. }\label{pair2}
\end{figure}

\begin{figure}[htbp]\centering 
\begin{center}
\includegraphics[width=5.8cm,height=5.8cm]{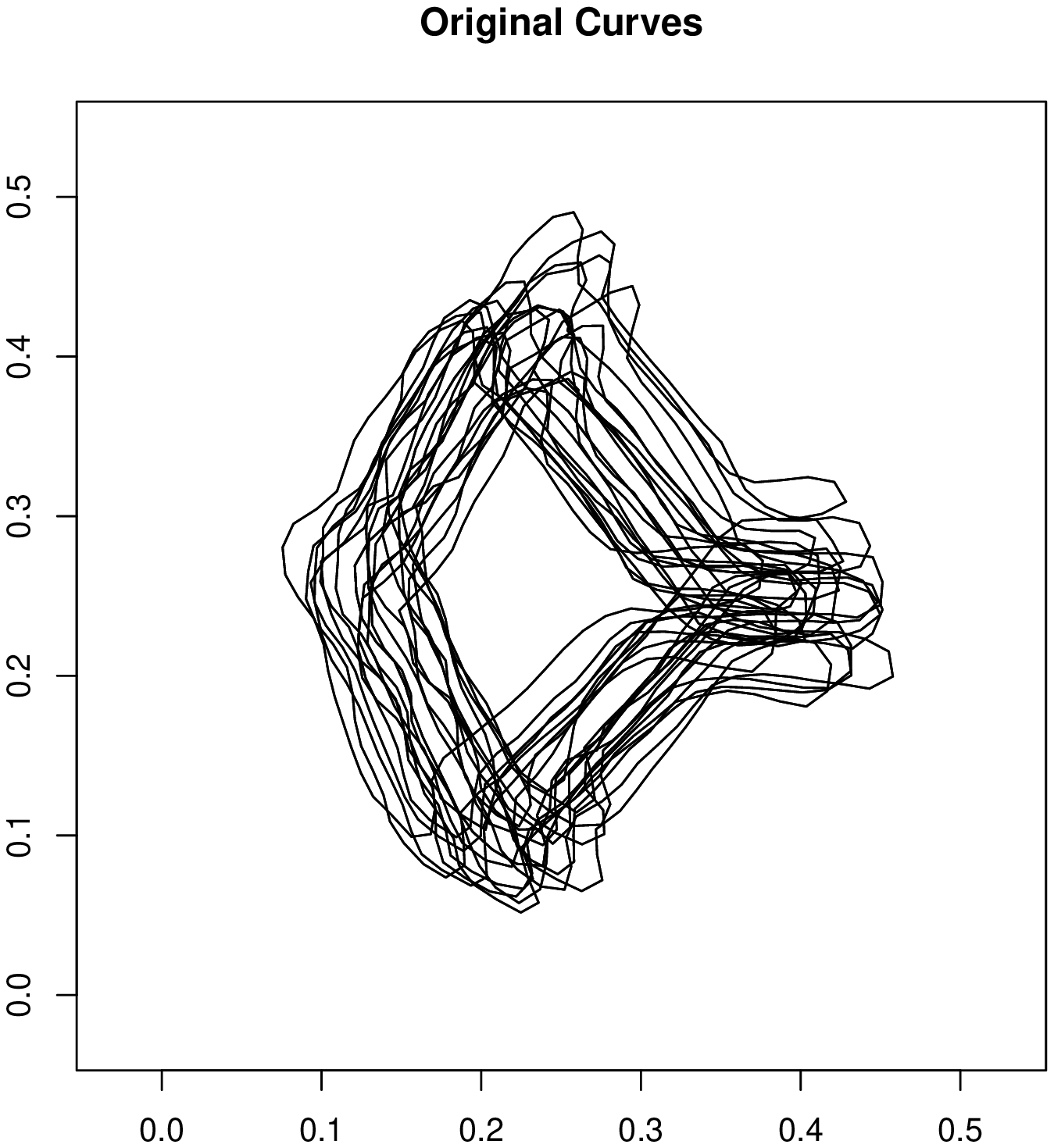}
\includegraphics[width=5.8cm,height=5.8cm]{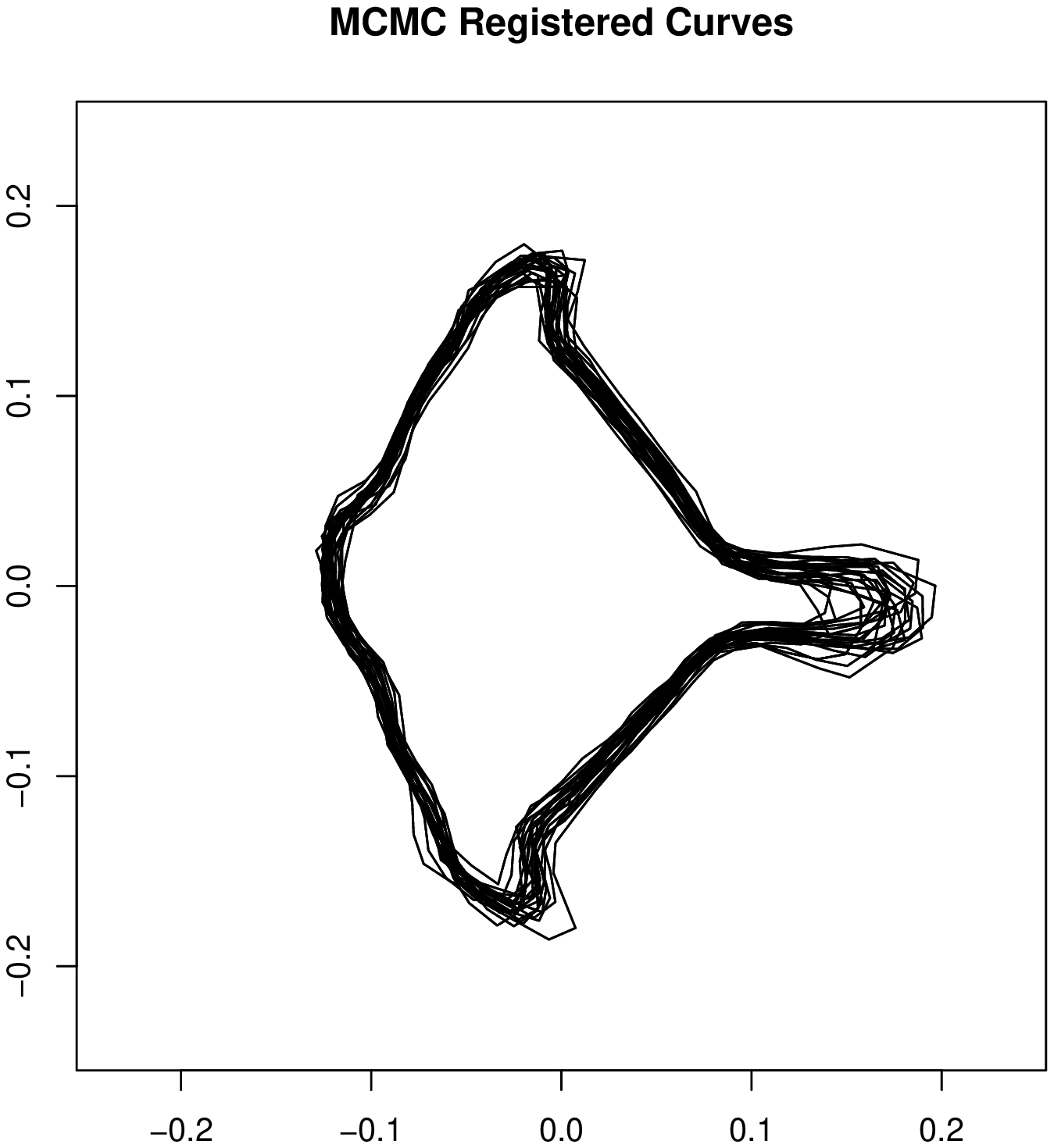}
\caption{The original curves from Small group, without and with registration.}\label{mult}
\end{center}
\end{figure}

We start our analysis by performing a pairwise comparison from the ambient space model, 
and we use the MCMC algorithm for pairwise matching with $50,000$ iterations. The $q$-functions are obtained by initial smoothing, 
and then normalized so that $\| q \|_2 = 1$. The registration is carried out using rotation through an angle $\theta$ about the origin, and 
a warping function $\gamma$.
 The original and registered pair (using a posterior mean) are shown in Figure \ref{pair}  
and the point-wise correspondence between the curves and a point-wise 95\% credibility interval for $\gamma(t)$ are shown in Figure \ref{pair2}.
The start point of the curve is fixed and is given by the left-most point on the curve in Figure \ref{pair2} that has a red line connecting the two bones. 
The narrower regions in the credibility interval correspond well with high curvature regions in the shapes. Convergence of the MCMC scheme 
was monitored by trace plots. 
We also applied the multiple curve registration, as shown in Figure \ref{mult}. 

In order to investigate the differences between the new Bayesian method and standard generalized Procrustes analysis on the 60 
landmarks we consider 
a classification study. 
For classification method A, the three 
group means are obtained through classical generalized Procrustes analysis \citep{Goodall91} using the {\tt shapes} package in R \citep{Dryden13} 
and each test curve is assigned to the trained group which is closest in terms of Procrustes distance. The Procrustes distance is calculated 
by minimizing the Euclidean sum of squares between the landmark configurations using translation, rotating and scale. 
For method B, the three group means are obtained using the posterior mean from the Bayesian model and each 
test curve is classified based on the elastic distance to the mean (i.e. using amplitude variability). 
For method C, all training dataset curves are registered in one pooled group using generalized Procrustes analysis 
and the Procrustes registered curves are used as the training data. Each test curve is aligned to the mean by 
ordinary Procrustes analysis. A Support Vector Machine (SVM) \citep{Changlin01} is then trained on the 
registered training curves and applied to the registered test curves. For method D, all training dataset curves are registered through the 
Bayesian model and their warped, registered versions are used as the training data. Each test curve is aligned 
to the mean by pairwise registration using MCMC. An SVM is then trained on the MCMC registered training curves and applied to 
the registered test curves.

A total of 100 Monte Carlo 
repetitions are run for each exercise, where the training data and test data are sampled from each group without replacement. In a single Monte Carlo 
repetition, 16 curves from the Small group, 20 from the Control group and 16 from the Large group 
(about two-thirds of the original data) are randomly selected as the training data, while the remaining 
24 curves are used as the test data. 
Method A gives an 80\% correct classification rate for the 
test data, and method B gives 78\% correct classification. In method C, the classification rate increases rate to 87\% while method 
D has the highest classification rate of 92\%. It is interesting to see that method A (Procrustes) does a little bit better than method B (Bayesian), 
although they are very similar.  We see an overall improvement in methods C and D compared to A and B. The main difference here is that methods 
C and D are using hyperplanes to classify between distributions for each group, rather than shape distances which are isotropic in nature. 
 Method D demonstrates the advantage in using the Bayesian MCMC method for registration with warping, rather than just using the equally spaced 
pseudo-landmarks with no warping.

\subsection{Mass spectrometry data}\label{massspec5.2}
Consider a one-dimension functional dataset of Total Ion Count (TIC) 
chromatograms of five individuals with acute myeloid leukemia, each with three replicates. 
The data were made available and described by \citet{Kochetal13} 
at the Mathematical Bioscience Institute (MBI), Columbus, Ohio, workshop on 
Statistics of Time Warpings and Phase Variations, November 2012.\footnote{{\tt http://mbi.osu.edu/2012/stwresources.html}}
After pre-processing, each of the 15 observations contains 2001 data points (intensities) from a truncated scan time of 20 to 220 minutes, 
with linear interpolation at the same time points of all 15 TIC curves. 
We carried out some further pre-processing including
baseline extraction (using cubic spline $\lambda = 5$) and smoothing for larger time points where excessive noise exists (with a cubic spline $\lambda = 0.4$).  
Some initial analysis at the workshop was carried out by \cite{Chengetal13proteins}, and 
there was a suggestion that the posterior exhibits signs of multimodality 
and that the Bayesian MCMC algorithm can become stuck in a local mode. In the analysis here we use simulated tempering from Section \ref{simtemp} and compare
the results to when using the algorithm without simulated tempering. 

We first consider a pairwise Bayesian registration between two TIC curves. The estimated warping function $\hat{\gamma}_A$ is taken as the posterior mean of MCMC samples after convergence and the prior for $\gamma$ is taken as ${\rm Dirichlet}(1)$, with  $M=40$ in the piecewise linear approximation.
In order to deal with the multimodality of the posterior, simulated tempering is employed \citep{Geyethom95} where ten 
levels of temperature are included and the initial phase for tuning to estimate the weights used in moving between levels is $50,000$ iterations. 
The two curves before and after alignment are given in Figure \ref{fig3} as seen earlier, and all the peaks look well aligned.
Convergence of the MCMC algorithm is monitored by traceplots of concentration parameter $\kappa$ and 
the log posterior, and there are no obvious violations of convergence (not shown).






%
%

When multiple curves are under consideration, a set of $(\hat{\gamma}_1,\dots,\hat{\gamma}_n)$ are taken from MCMC samples. 
Again the convergence of the MCMC algorithm is checked by the traceplots of the concentration parameter $\kappa$ and log posterior, which seem fine for these data (not shown).
For this particular dataset 14 spiked proteins in the data have been identified in each scan by 
the experimenters, which can be regarded as an answer key. If the  registration results agree with the answer key, then 
the positions of the 14 spiked proteins should coincide after registration. 

\begin{figure}[htbp]\centering 

\begin{tabular}{c}
\includegraphics[height=6cm]{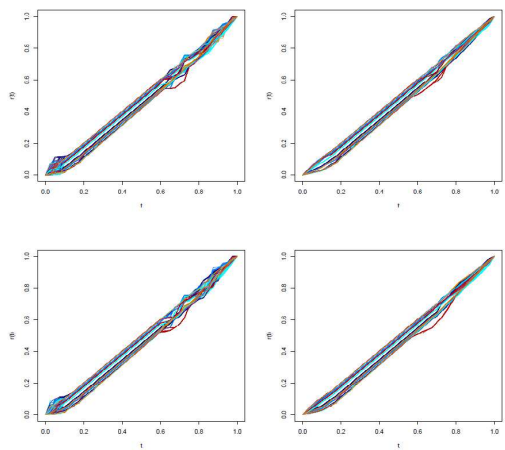} 
\end{tabular}
\caption{Samples of groups of warping functions when $\gamma\sim {\rm Dirichlet}(1)$ and $\gamma\sim {\rm Dirichlet}(100)$. The top and bottom
rows indicate the results without and with simulated tempering respectively.} \label{warps}
\end{figure}
\begin{figure}[htbp]\centering 
\begin{tabular}{c}
\includegraphics[height=6cm,width=6cm]{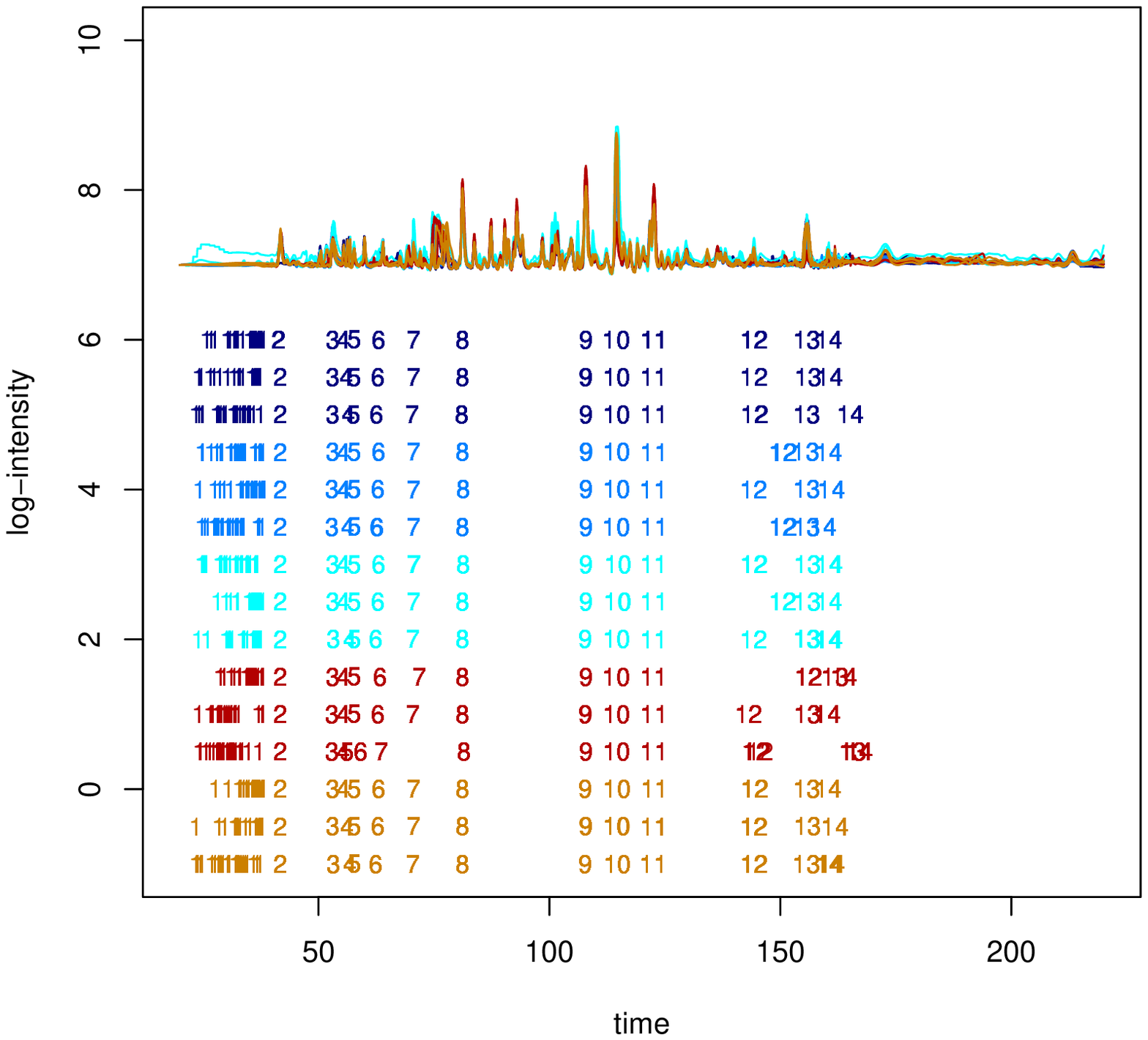} 
\includegraphics[height=6cm,width=6cm]{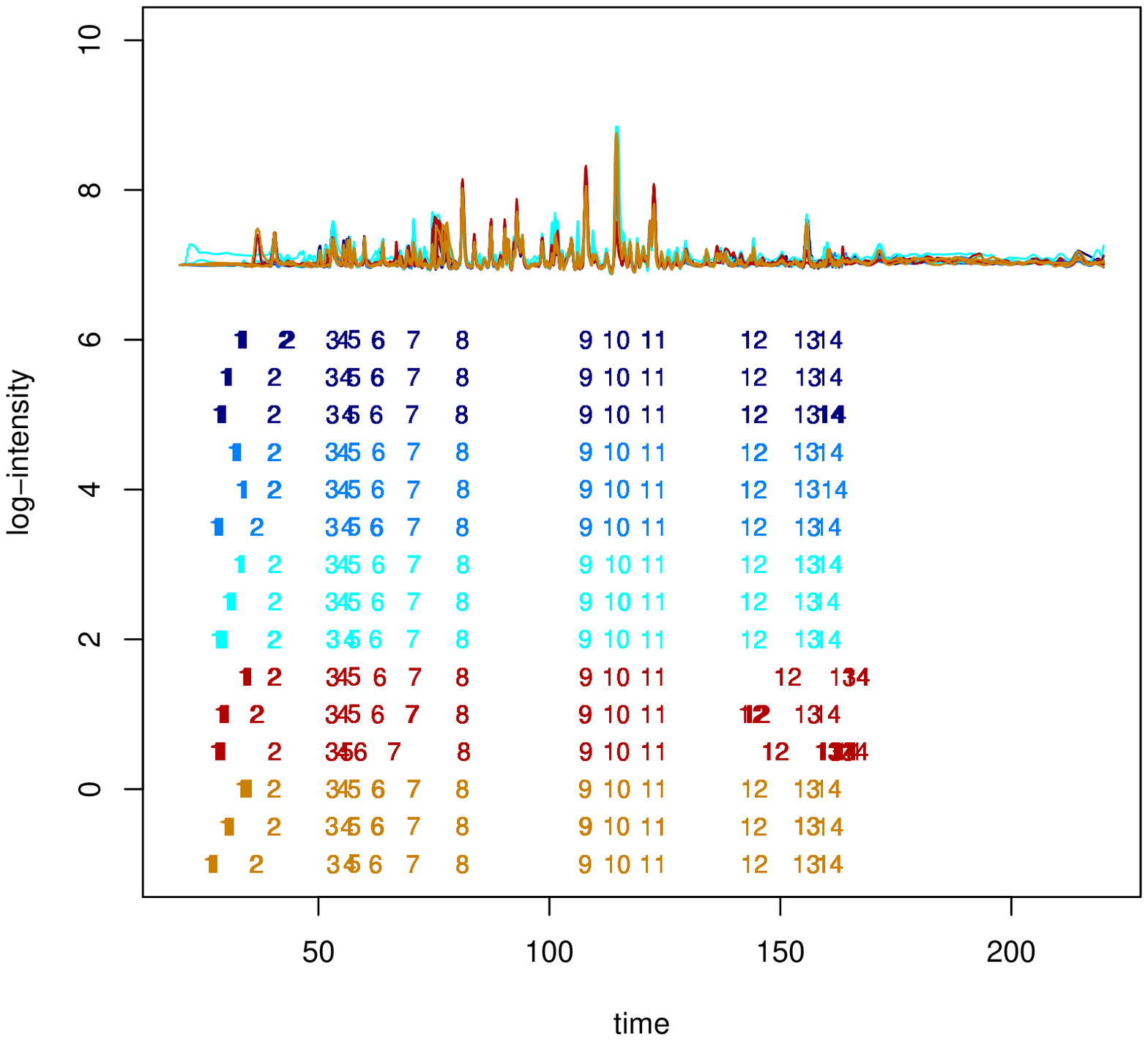}\\
\includegraphics[height=6cm,width=6cm]{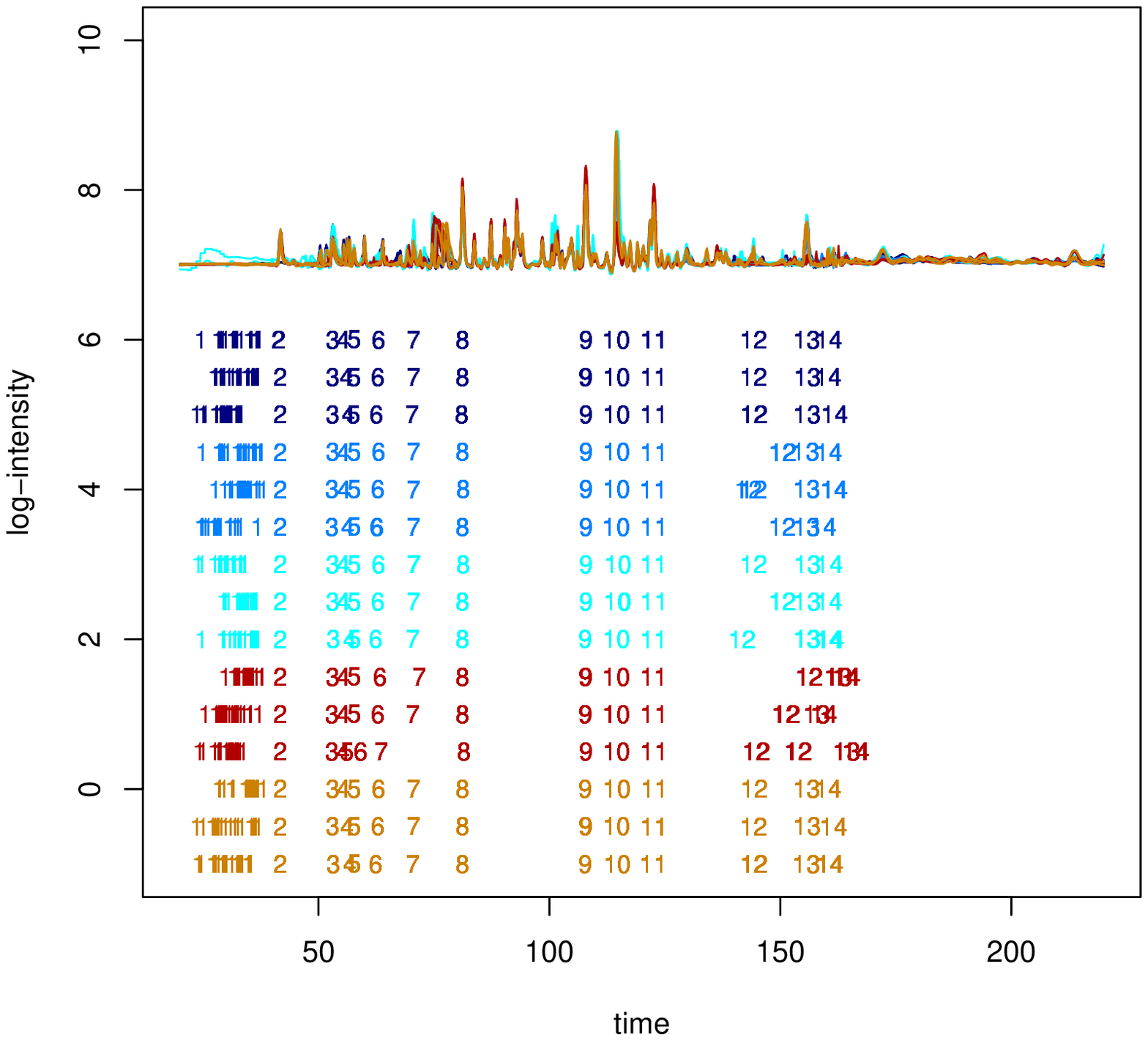} 
\includegraphics[height=6cm,width=6cm]{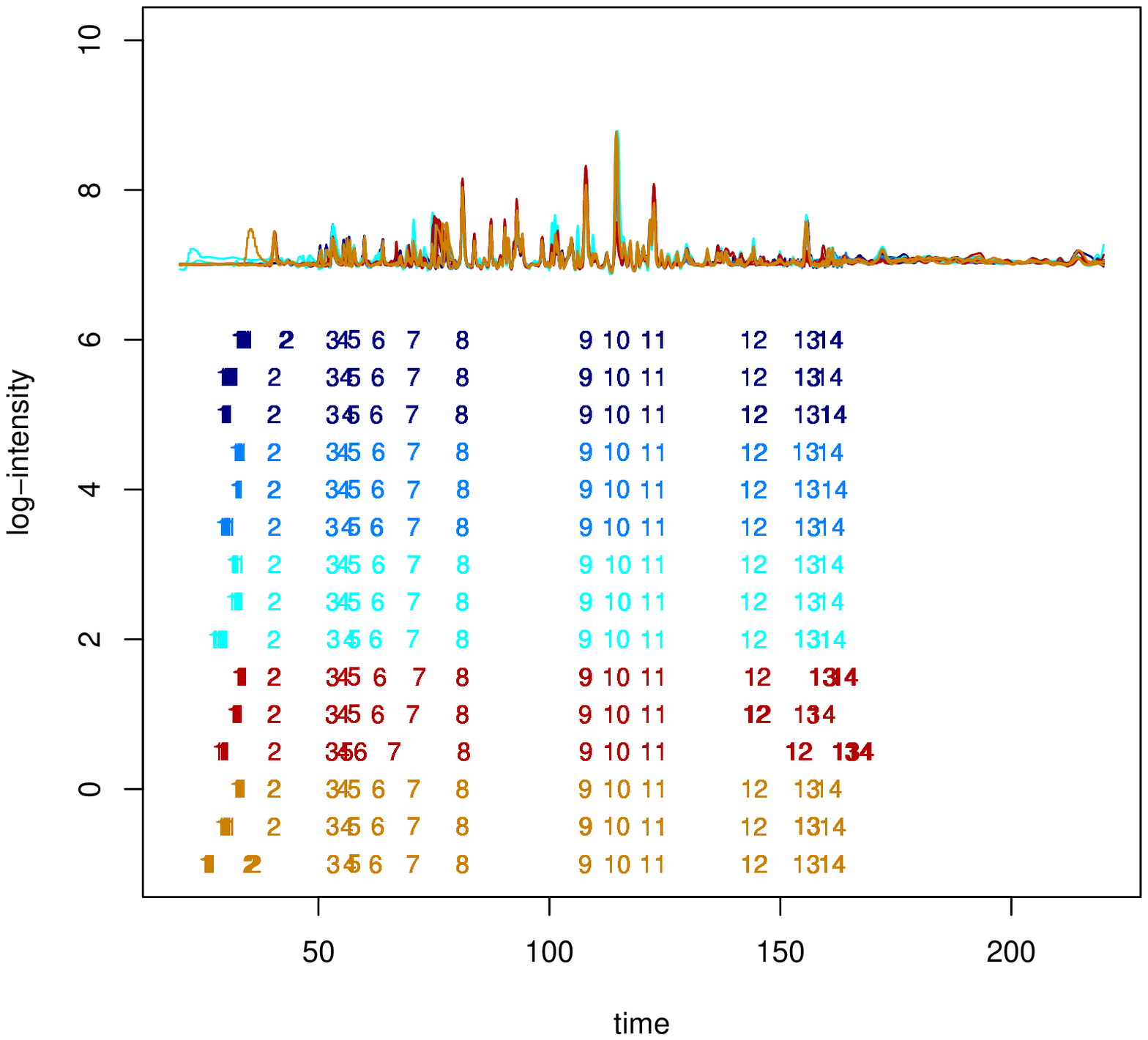} 
\end{tabular}
\caption{Multiple registration results based on samples of $(\gamma_1,\gamma_2,\dots,\gamma_{n})$ when $\gamma\sim {\rm Dirichlet}(1)$ (left) and $\gamma\sim {\rm Dirichlet}(100)$ (right). The top and bottom rows show the results without and with simulated tempering, respectively.} \label{mcmcreg}
\end{figure}

To indicate the posterior variability of warping functions under different priors, 
realizations of the warping functions are taken from the MCMC simulation  and are shown in Figure \ref{warps}. 
The registration results based on these samples of warping functions are also shown in Figure \ref{mcmcreg}. 
The first row of integers corresponds to
the warped spike positions for individual 1, replicate 1. The second row corresponds to individual 1, replicate 2, etc. 
In the ideal scenario of perfect alignment we should have all sets of numbers in 14 vertical columns. 
In this analysis the MCMC algorithm was run for 
50,000 iterations (after tuning the weights for simulated tempering) and we display every 1000th value after the burn-in period of 25,000
iterations, i.e. each number is shown 25 times. 
From Figure \ref{mcmcreg}, we see that the main variability when $\gamma\sim {\rm Dirichlet}(1)$ lies in the position of 1, where the curve is flatter and thus contains less information, while when $\gamma\sim {\rm Dirichlet}(100)$ the variability is so small that the different numbers are only slightly different, 
indicating a very tight posterior distribution. We see that the registration results under both priors look reasonably good as most positions line up in a vertical line. 
However, we do notice that the stronger prior clearly helps the alignment at positions 1, 12, 13 and 14. 

From Figure \ref{mcmcreg} the use of simulated tempering provides 
an improvement in the MCMC 
algorithm compared to not using it, where there is a danger of becoming stuck in local modes. Using simulated tempering and 
$a=100$ all but two spike 2's and all but one of the spike 12's are well aligned. 
Since the data mainly exhibit translational effects for registration, the strong prior ($a=100$) is particularly appropriate here. 

\section{Discussion}
In this paper we state the distinctions between three spaces of interest: the original, ambient and quotient space. 
We compare the ambient space estimator and quotient space estimator in simulation studies, 
and explain the similarity in certain situations through a Laplace approximation. An important 
component is that we incorporate prior information 
about the amount of warping, which is particularly useful in the mass spectrometry application, where too much warping is not desirable. Naturally the choice of prior 
is important and will of course be problem specific, however in the mass spectrometry data it was clear that translations are particularly important, and our prior 
is weighted strongly towards this feature.

Note that for matching between two functions we also can use multiple alignment, which also involves estimating 
the mean function, instead of the pairwise method. Although the multiple alignment method
appears to be a less efficient approach due to the need for the mean function as parameters, 
it does have the property that the prior would be invariant under a common reparameterization of both curves.

Although we have focused on 1D and 2D applications the Bayesian methodology can be extended to higher dimensions, for example analysing the shape of 
3D surface shapes using the square root normal fields \citep{Jermetal12}.

\bibliographystyle{apalike}
\bibliography{/maths/staff/pmzild/tex/bibtex/fullref}

\end{document}